\documentclass[floatfix,twocolumn,showpacs,aps,tightenlines]{revtex4-1}
\usepackage{graphicx}
\usepackage{color}
\usepackage{psfrag}
\usepackage{dcolumn}% Align table columns on decimal point
\usepackage{bm}% bold math

\usepackage{amssymb}
\usepackage{amsmath}
\usepackage{amsfonts}
\usepackage{longtable}
\usepackage{xspace}
\usepackage[export]{adjustbox}

\newcommand{\Spar}{\tilde S_\|}
\newcommand{\Dpar}{\tilde \Delta_\|}
\newcommand{\dmt}{\delta{m}}

\newcommand{\beq}{\begin{equation}}
\newcommand{\eeq}{\end{equation}}

\newcommand{\be}{\begin{equation}}
\newcommand{\ee}{\end{equation}}
\newcommand{\bea}{\begin{eqnarray}}
\newcommand{\eea}{\end{eqnarray}}
\newcommand{\bes}{\begin{subequations}}
\newcommand{\ees}{\end{subequations}}

\usepackage{bm}

\begin{document}

\title{The hangup effect in unequal mass binary black hole mergers
and further studies of their gravitational radiation and remnant properties}

\author{James Healy}
\author{Carlos O. Lousto}
\affiliation{Center for Computational Relativity and Gravitation,
School of Mathematical Sciences,
Rochester Institute of Technology, 85 Lomb Memorial Drive, Rochester,
New York 14623}

\date{\today}

\begin{abstract}
We present the results of 74 new simulations of nonprecessing spinning
black hole binaries with mass ratios $q=m_1/m_2$ in the range
$1/7\leq q\leq1$ and individual spins covering the parameter space
$-0.95\leq\alpha_{1,2}\leq0.95$ with one runs with spins of $\pm0.95$.
We supplement those runs with 107 previous
simulations to study the hangup effect in black hole mergers,
i.e. the delay or prompt merger of spinning holes with respect to non
spinning binaries. 
We perform the numerical evolution for typically the last
ten orbits before the merger and down to the formation of the final remnant
black hole. This allows us to study the hangup effect for unequal mass
binaries leading us to identify the spin variable that controls
the number of orbits before merger as $\vec{S}_{hu}\cdot{\hat{L}},$ 
where $\vec{S}_{hu}=(1+\frac12\frac{m_2}{m_1})\vec{S}_1+(1+\frac12\frac{m_1}{m_2})\vec{S}_2$.
We also combine the total results of those 181 simulations to obtain improved fitting formulae for the remnant final black hole mass, spin and recoil velocity as well as for the peak luminosity and peak frequency of the gravitational strain, and find new correlations among them.
This accurate new set of simulations enhances the number of available numerical relativity waveforms available for parameter estimation of gravitational wave observations.

\end{abstract}

 \maketitle

\section{Introduction}\label{sec:Intro}

The breakthroughs~\cite{Pretorius:2005gq,Campanelli:2005dd,Baker:2005vv}
in numerical relativity enabled the detailed predictions for the
gravitational waves from the late inspiral, plunge, merger and
ringdown of black hole binary systems (BHB). The first generic, long-term
precessing binary black hole evolution without any symmetry have been 
performed a decade ago in 
Ref.~\cite{Campanelli:2008nk}, where a detailed comparison with 
post-Newtonian $\ell=2,3$ waveforms was made. 
Gravitational waves from the merger of black holes have been now 
directly observed by LIGO: GW150914\cite{Abbott:2016blz} and 
GW151226\cite{Abbott:2016nmj} during the first observing run O1\cite{TheLIGOScientific:2016pea}, and GW170104 \cite{Abbott:2017vtc}, GW170608 \cite{Abbott:2017gyy}, and GW170814 (jointly with Virgo) \cite{Abbott:2017oio} during
the second observing run, O2.
Direct comparison of targeted full numerical simulations with the first events of the observing run have been performed in \cite{Abbott:2016apu} for GW150914
(with \cite{Lovelace:2016uwp} providing the details of the simulation displayed
in Fig. 1 of \cite{Abbott:2016blz}) and in \cite{Healy:2017abq} for GW170104.

Numerical relativity techniques allow us to explore the late
binary dynamics, beyond the post-Newtonian regime. Notable 
early examples  are, for instance, the study of the {\it hangup} effect, 
i.e. the role individual black hole
spins play to delay or accelerate their merger \cite{Campanelli:2006uy},
and the determination of the magnitude and direction 
of the potentially large (up to 5000km/s) {\it recoil} velocity of the 
final merged black hole
\cite{Campanelli:2007ew,Campanelli:2007cga,Lousto:2011kp},
and the effects of precession, such as the
{\it flip-flop} of individual spins during the orbital phase
\cite{Lousto:2014ida,Lousto:2015uwa,Lousto:2016nlp}.

In Refs.~\cite{Healy:2014yta} and ~\cite{Healy:2016lce}
we used 37 plus 71 original runs (and those available in 
the literature) to determine fitting formulae that relate aligned
spin binaries orbital parameters $(q,\alpha_1,\alpha_2)$ to
the final black hole mass, spin and recoil
$(m_f,\alpha_f,V_f)$. Here we revisit this scenario and extend the
study to investigate the hangup effect for unequal mass, 
nonprecessing binaries.

The paper is organized as follows. Next Section \ref{sec:FN}
describe the methods and criteria for producing the new
simulations. In Sec.~\ref{sec:Hangup} we review
the characterization of the hangup effect for numerical and post-Newtonian
approaches. We set up new simulations of
unequal mass binaries in Sec.~\ref{sec:Simulations} to find an effective
spin description of the hangup. In Section \ref{sec:Omegapeak}
we model the peak luminosity from the gravitational wave strain and its frequency as a function of the parameters of the precursor binary. In Sec.
\ref{sec:Remnant} we use the new data to improve the remnant black hole
mass, spin and recoil velocity fits. Sec.\ref{sec:Correlations} discusses
correlations among the above quantities as directly obtained from the
full set of 181 simulations. We
conclude with a discussion in Sec.~\ref{sec:Discussion} of the use 
of these results in the modeling of gravitational waves and its 
potential extensions to precessing binaries.

\section{Full Numerical Evolutions}\label{sec:FN}

We evolve the following BBH data sets using the {\sc
LazEv}~\cite{Zlochower:2005bj} implementation of the moving puncture
approach~\cite{Campanelli:2005dd,Baker:2005vv} with the conformal
function $W=\sqrt{\chi}=\exp(-2\phi)$ suggested by
Ref.~\cite{Marronetti:2007wz}.  For the run presented here, we use
centered, sixth-order finite differencing in
space~\cite{Lousto:2007rj} and a fourth-order Runge Kutta time
integrator (Note that we do not upwind the advection terms.)
and a 7th-order Kreiss-Oliger dissipation operator.
This sixth-order spatial finite difference allow us to gain
a factor $\sim4/3$ with the respect to the eight-order implementation
due to the reduction of the ghost zones from 4 to 3.
We also allowed for a Courant factor $CFL=1/3$ instead of the
previous $CFL=1/4$ \cite{Zlochower:2012fk} 
gaining another speedup factor of 4/3. 
We verified that for this relaxing of the time integration step
we still conserve the horizon masses and spins of the individual
black holes during evolution and the phase of the gravitational
waveforms to acceptable levels. This plus the use of the new
Xsede supercomputer {\it Comet} at 
SDSC~\footnote{https://portal.xsede.org/sdsc-comet} lead to typical
evolution speeds of $250M/day$ on 16 nodes, 
where M is the mass that defines the scale of the 
simulation. Note that our
previous \cite{Lousto:2013oza,Lousto:2015uwa}
comparable simulations averages $\sim100M/day$.

Our code uses the {\sc EinsteinToolkit}~\cite{Loffler:2011ay,
einsteintoolkit} / {\sc Cactus}~\cite{cactus_web} /
{\sc Carpet}~\cite{Schnetter-etal-03b}
infrastructure.  The {\sc
Carpet} mesh refinement driver provides a
``moving boxes'' style of mesh refinement. In this approach, refined
grids of fixed size are arranged about the coordinate centers of both
holes.  The {\sc Carpet} code then moves these fine grids about the
computational domain by following the trajectories of the two BHs.

We use {\sc AHFinderDirect}~\cite{Thornburg2003:AH-finding} to locate
apparent horizons.  We measure the magnitude of the horizon spin using
the {\it isolated horizon} (IH) algorithm detailed in
Ref.~\cite{Dreyer02a} and as implemented in Ref.~\cite{Campanelli:2006fy}.
Note that once we have the horizon spin, we can calculate the horizon
mass via the Christodoulou formula 
%\begin{equation}
${m_H} = \sqrt{m_{\rm irr}^2 + S_H^2/(4 m_{\rm irr}^2)}\,,$
%\end{equation}
where $m_{\rm irr} = \sqrt{A/(16 \pi)}$, $A$ is the surface area of
the horizon, and $S_H$ is the spin angular momentum of the BH.
%(in units of $m^2$).
In the tables below, we use the variation in the
measured horizon irreducible mass and spin during the simulation as a
measure of the error in computing these quantities.  
We measure radiated energy,
linear momentum, and angular momentum, in terms of the radiative Weyl
Scalar $\psi_4$, using the formulas provided in
Refs.~\cite{Campanelli:1998jv,Lousto:2007mh}. However, rather than
using the full $\psi_4$, we decompose it into $\ell$ and $\tilde{m}$ modes and
solve for the radiated linear momentum, dropping terms with $\ell >
6$.  The formulas in Refs.~\cite{Campanelli:1998jv,Lousto:2007mh} are
valid at $r=\infty$.  We extract the radiated energy-momentum at
finite radius and extrapolate to $r=\infty$. We find that the new
perturbative extrapolation described in Ref.~\cite{Nakano:2015pta} provides the
most accurate waveforms. While the difference of fitting both linear and
quadratic extrapolations provides an independent measure of the error.

\begin{figure}
  \includegraphics[angle=270,width=0.88\columnwidth,left]{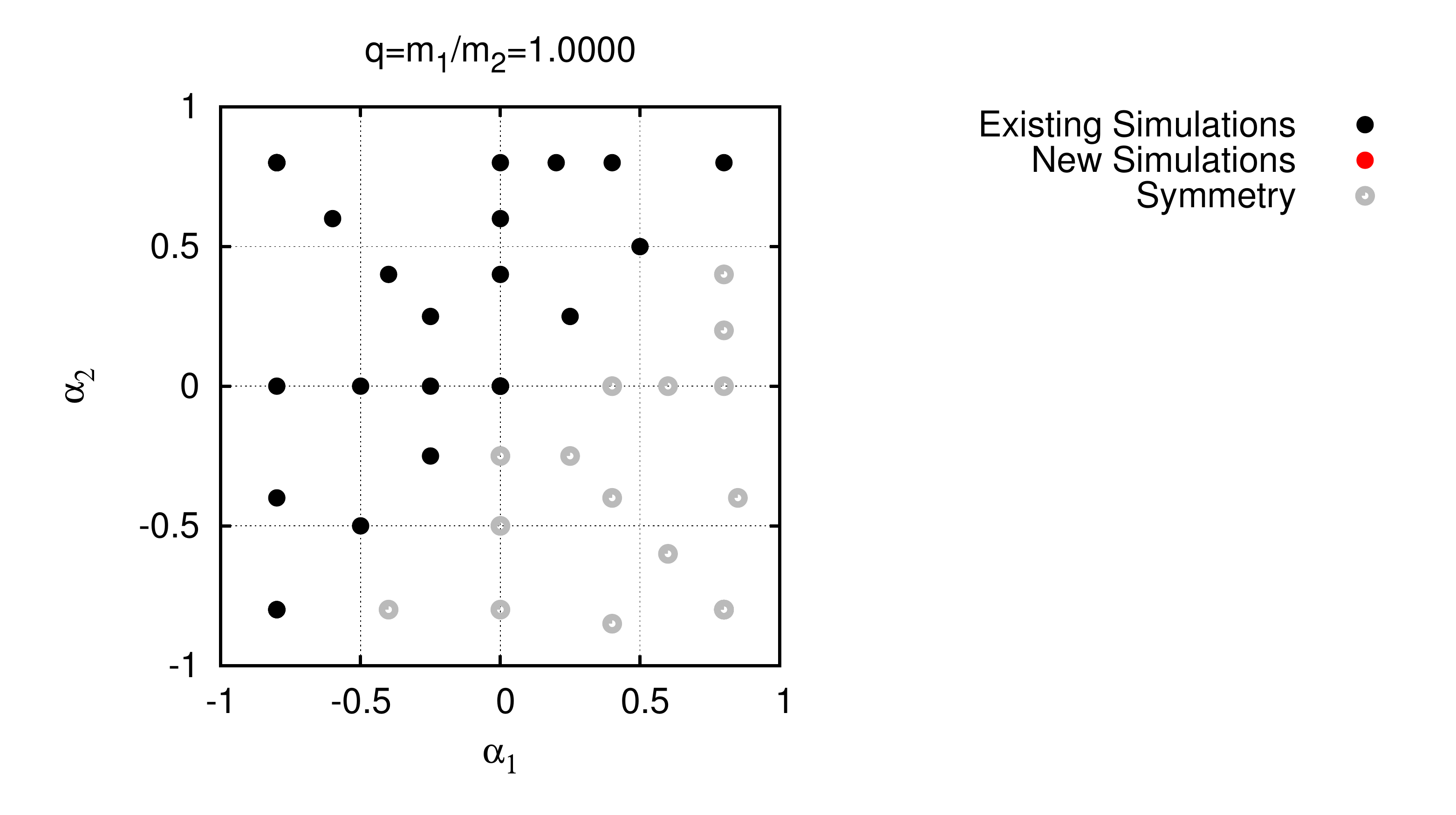}\\
  \includegraphics[angle=270,width=0.49\columnwidth]{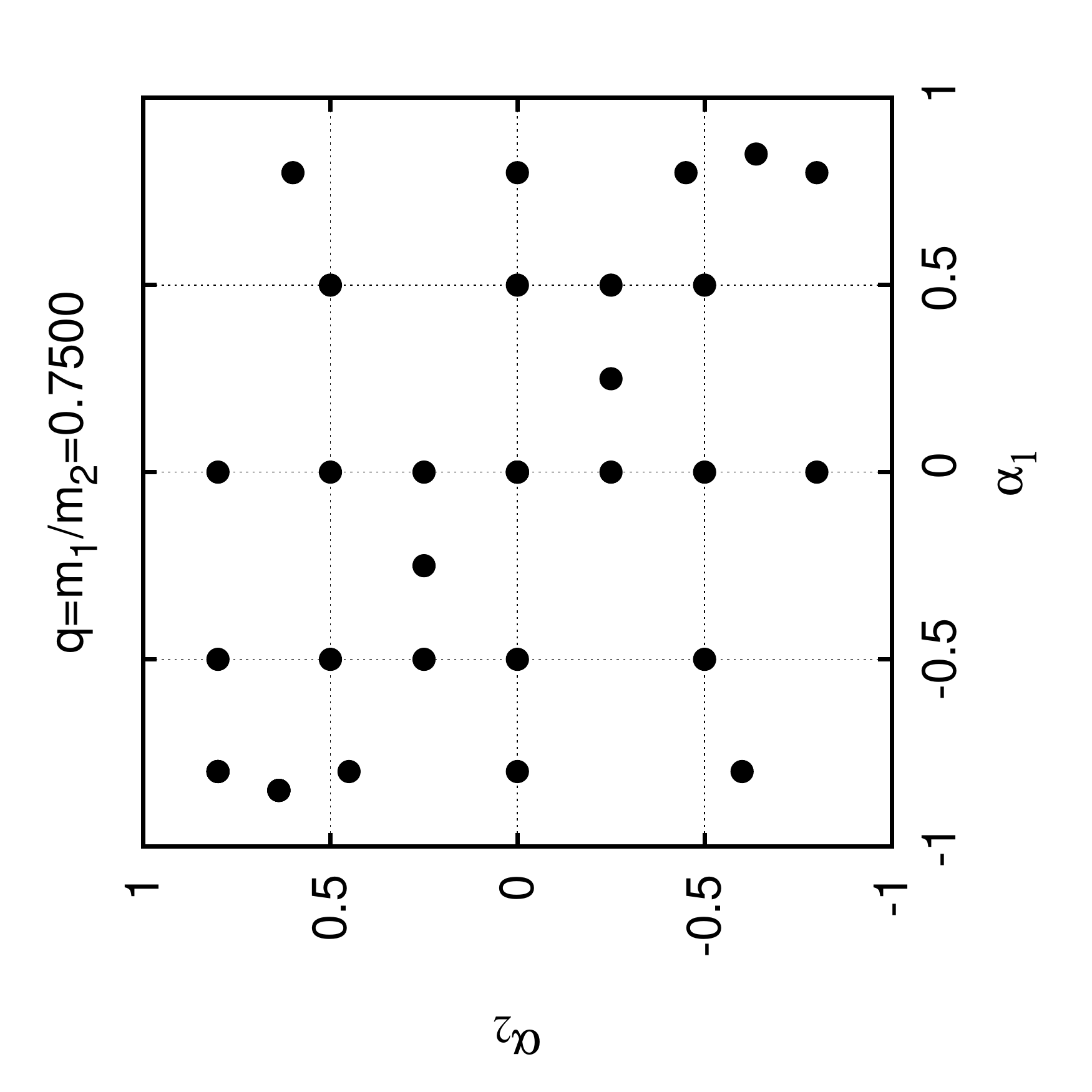}
  \includegraphics[angle=270,width=0.49\columnwidth]{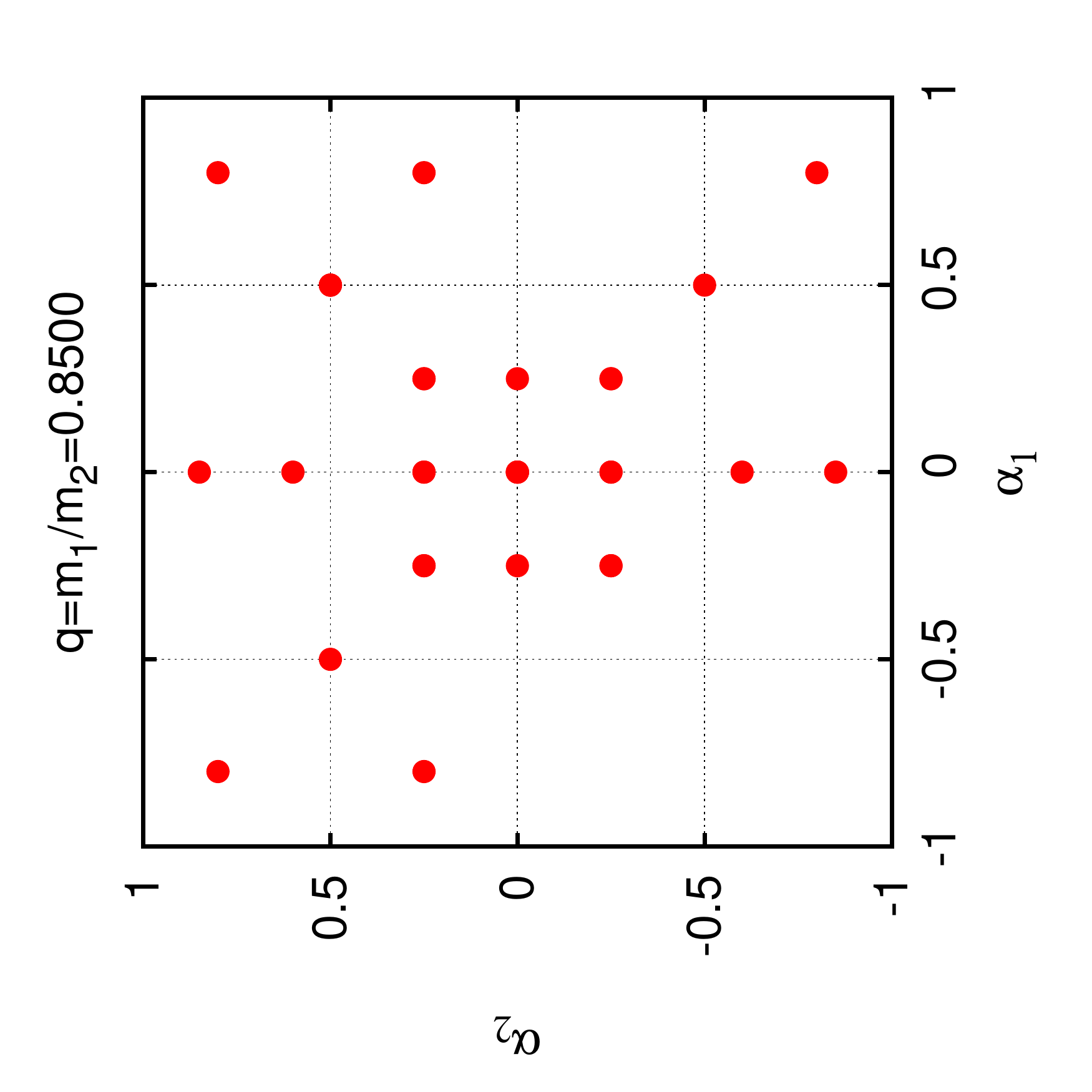}\\
  \includegraphics[angle=270,width=0.49\columnwidth]{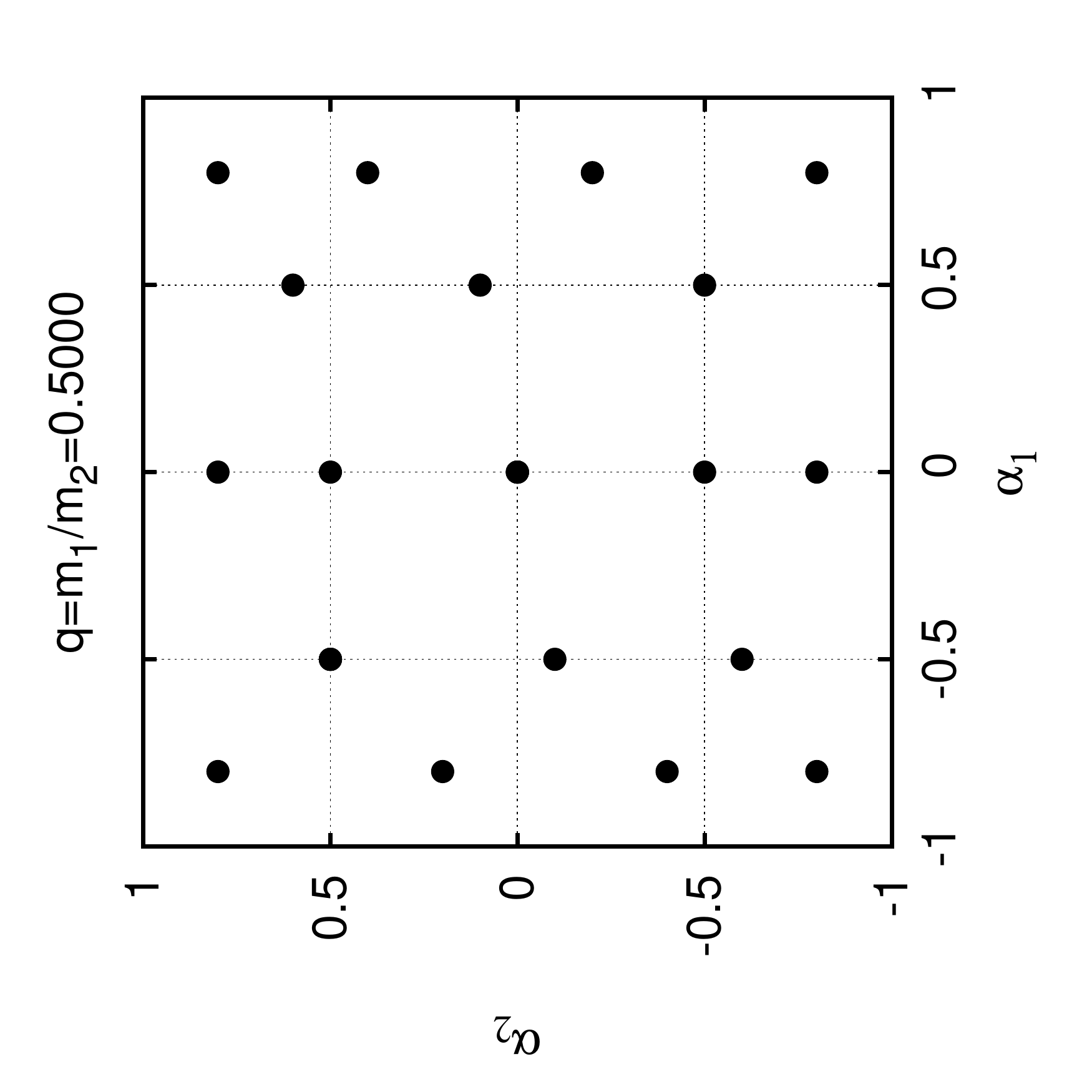}
  \includegraphics[angle=270,width=0.49\columnwidth]{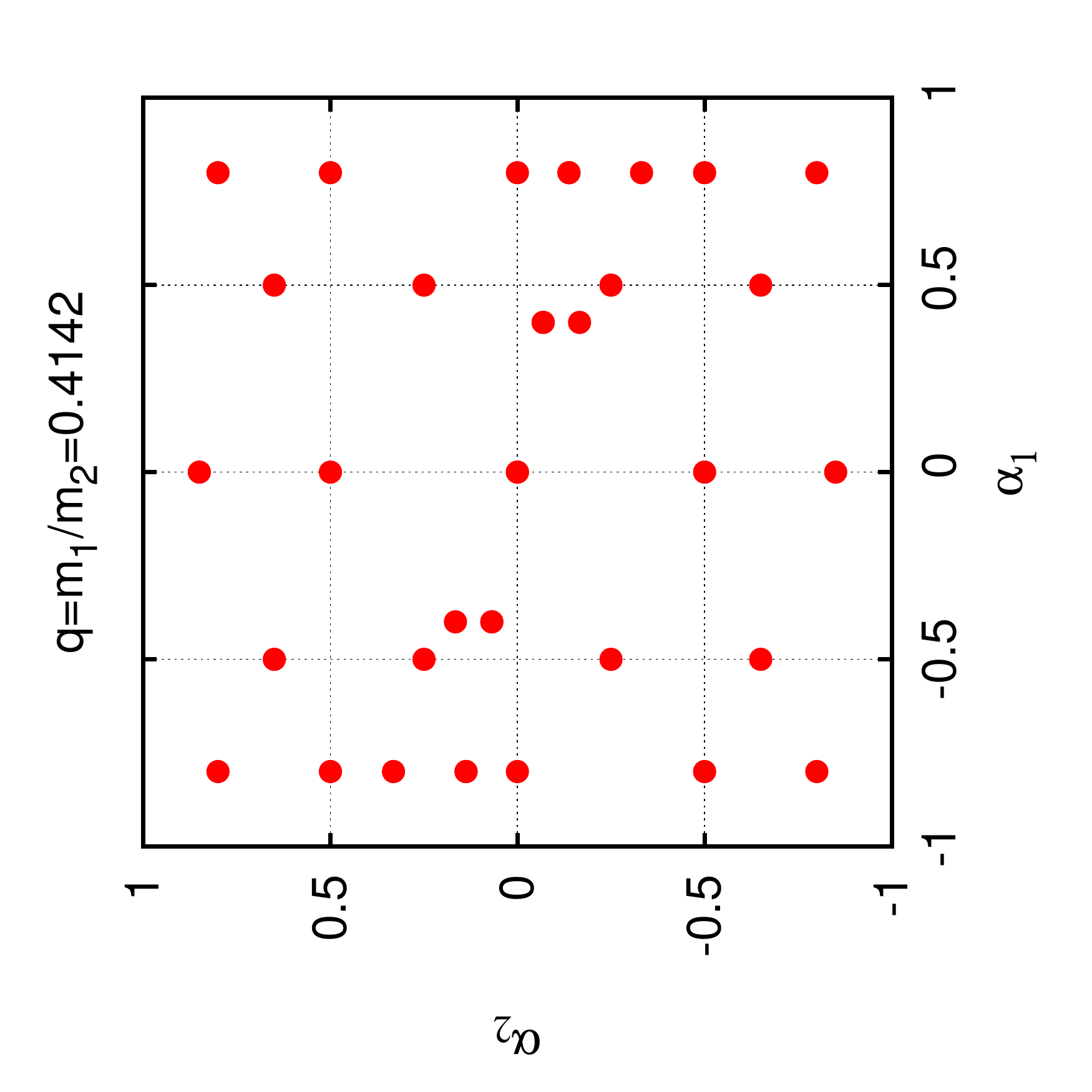}\\
  \includegraphics[angle=270,width=0.49\columnwidth]{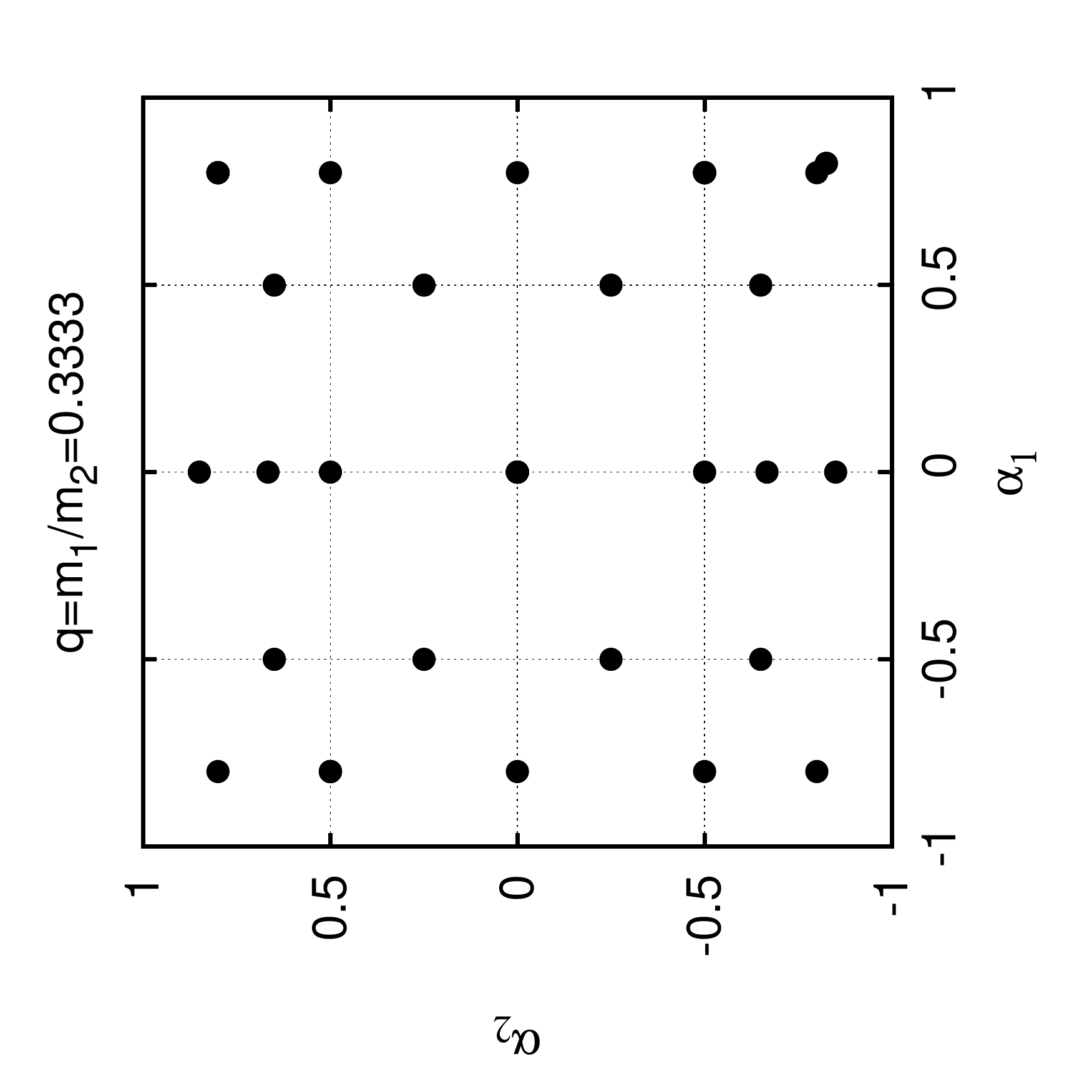}
  \includegraphics[angle=270,width=0.49\columnwidth]{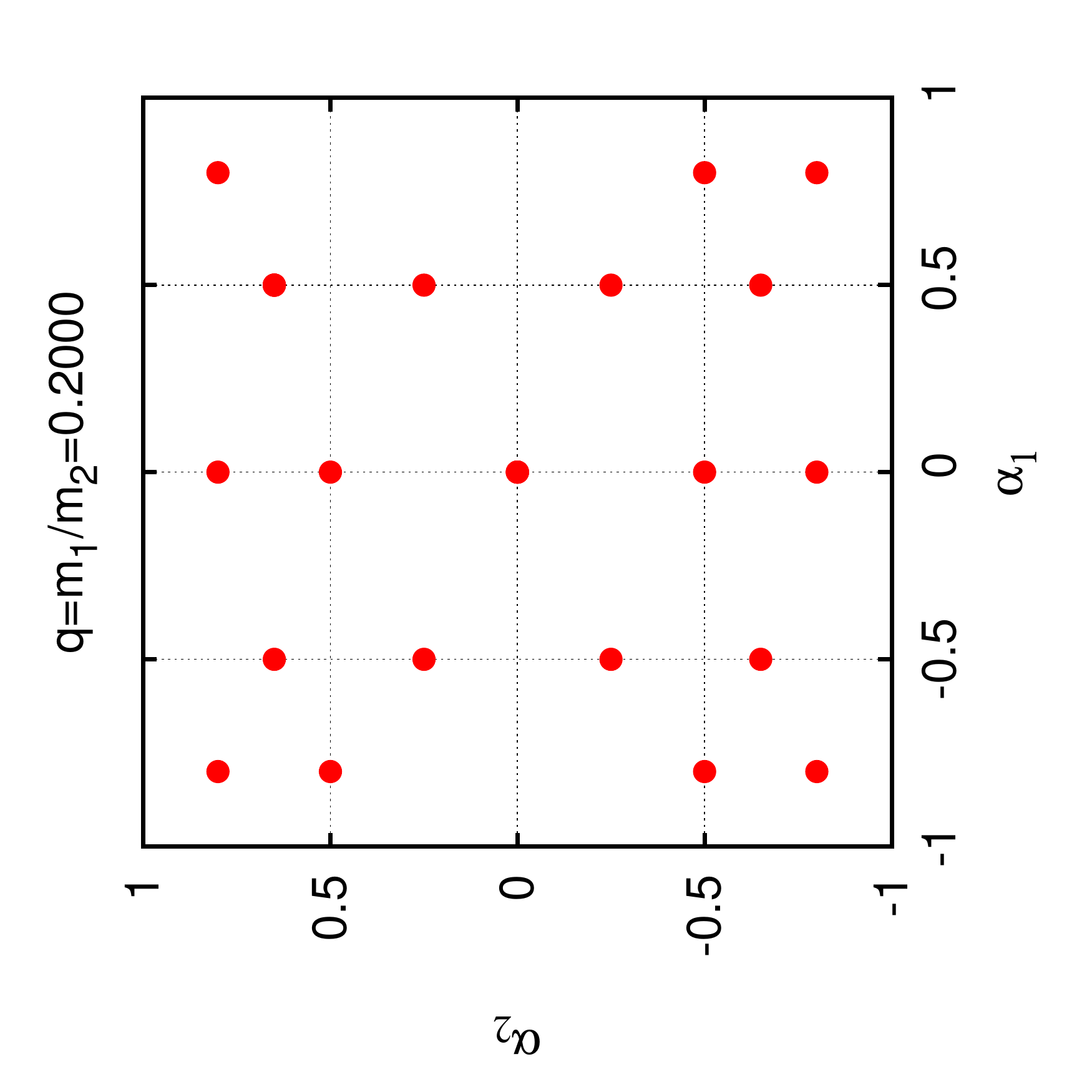}
  \caption{And overview of the spin and mass ratio parameters of 
    the new simulations presented in this paper. Labeled in red are 
68 new runs for mass ratios $q=0.85,0.4142,0.20$.  6 additional
    runs were included in the analysis not shown in these figures.
    The black points for mass ratios $q=1.00,0.75,0.50,0.33$
are runs completed for previous studies.
\label{fig:runs}}
\end{figure}

In this paper we have performed a new set of aligned spin BBH simulations 
targeted at supplementing the existing ones toward completion of a data
bank covering comparable BBH mass ratios down to 1:5.
Figure~\ref{fig:runs} gives an overview of the new regions of parameter
space covered in red (68 simulations)
and the coverage of our previous studies in black (107). 
6 additional simulations, designed for individual targeted studies are
also included. The total of 181 simulations are used for the
hangup studies of unequal mass, nonprecessing, binaries described 
in this paper.

Table \ref{tab:ID} including the initial data for all the new 74 
simulations is provided in the appendix \ref{app:ID}. We also provide the 
values of the individual masses and spins once they settle to 
equilibrium values from the initial data radiation content in 
table \ref{tab:IDr} as they provide a more physical
reference value.

\section{Hangup}\label{sec:Hangup}

The hangup effect is the strongest dynamical effect spins of individual
black holes have on their late time binary inspiral and merger~\cite{Campanelli:2006uy}.
This effect delays or prompts the merger speed 
of black hole binaries according to the sign of the spin-orbit coupling
$\vec{S}\cdot\vec{L}$, as this term has an additional repulsive 
(attractive) pull that is larger (smaller) than zero \cite{Kidder:1995zr}.
The strength of the effect on the merger process was first evaluated 
in \cite{Campanelli:2006uy} through full numerical simulations and 
was found to be much larger than expected from
the post-Newtonian analysis. Follow up work confirmed the 
strength of the hangup effect up to very large spins 
\cite{Hannam:2007wf,Hemberger:2013hsa}.

The original work~\cite{Campanelli:2006uy} was performed for equal mass, 
equal (anti-)aligned spins with the orbital angular momentum binaries.
The hangup effect, as later shown in \cite{Lousto:2013wta},
continues to be the most important effect in equal mass precessing binaries.
Here we build on the data bank of simulations for (anti-)aligned spins binaries
described in \cite{Healy:2014yta,Healy:2016lce,Healy:2017psd} and 
supplement it with 74 new simulations to analyze their dynamics in 
detail and determine what is the spin and mass ratio variables dependence
that controls the hangup effect in the unequal mass binaries cases.

In addition to those aligned runs,
here we also explore the interesting case of the possibility of having
a residual hangup effect even if the total spin of the binary is
zero. We would like to verify this directly on full nonlinear simulations
of binary black holes independently of the PN expansions. Note that
the assumption that the vanishing addition of the spins 
$\vec{S}=\vec{S}_1+\vec{S}_2=0$ leads to no effects has been used 
in developing some early models of the remnant formulae~\cite{Rezzolla:2007rz}.

In what follows we will use the following notation
(the tilde over variables denote the dimensionless normalization by $1/m^2$)
\begin{eqnarray}\label{eq:notation}
m &&= m_1 + m_2,\quad \delta m = \frac{m_1 - m_2}{m},\\
\tilde{\vec{S}}&& = (\vec S_1 + \vec S_2)/m^2=(\vec\alpha_2+q^2\vec\alpha_1)/(1+q)^2,\\
\tilde{\vec{\Delta}}&&=(\vec S_2/m_2 - \vec S_1/m_1)/m
=(\vec\alpha_2-q\vec\alpha_1)/(1+q),
\end{eqnarray}
where $m_i$ is the mass of BH $i=1,2$ and $\vec S_i$ is the spin of BH
$i$.
We also use the auxiliary variables
\beq
 \eta = \frac{m_1 m_2}{m^2},\quad q=\frac{m_1}{m_2},\quad
 \vec \alpha_i = \vec S_i/m_i^2,
\eeq
where $|\vec \alpha_i| \leq 1$ is the dimensionless spin of BH $i$,
and we use the convention that $m_1 \leq m_2$ and hence $q\leq 1$.
Here the index $\perp$ and $\|$
refer to components perpendicular to and parallel to
 the orbital angular momentum $\vec{L}$. We also define unit vectors
using ``hat'' labels, for instance as in $\hat{L}$.

There are two candidate effective spin parameters, $\vec{S}_0$ and $\vec{S}_{eff}$, 
that we can use to describe the hangup effects
(the number of orbits to merger from a fiducial initial orbital
frequency relative to the nonspinning case,
as studied in the original work ~\cite{Campanelli:2006uy}).
They come from the 2PN Hamiltonian spin dynamics
\cite{Damour:2001tu,Kidder:1995zr}, where
\begin{eqnarray}\label{eq:S0Seff}
\frac12S_{0}&=&\left(\vec{S}\cdot\hat{L}+\frac{1}{2}\,\delta{m}\vec{\Delta}\cdot\hat{L}\right),\\
\frac47S_{eff}&=&\left(\vec{S}\cdot\hat{L}+\frac{3}{7}\,\delta{m}\vec{\Delta}\cdot\hat{L}\right),
\end{eqnarray}
where we will normalize effective spins to produce $\vec{S}_2$, the large
black hole spin, in the extreme mass ratio limit.

%%%%%%%
A third candidate and a more explicit computation of the hangup effect can be
derived from 
Kidder's \cite{Kidder:1995zr} Eq. (4.16) that calculates the
accumulated orbital phase of the binary from 
the evolution of the orbital frequency
\begin{equation}
\Psi \equiv \int^{t_f}_{t_i} \omega dt = \int^{\omega_f}_{\omega_i}
{\omega \over \dot \omega} d\omega ,
\end{equation}
where $t_i$ is the initial time considered
(corresponding to a lower frequency $\omega_i$) and
$t_f$ is the final time at which the merger occurs
(corresponding to an upper frequency $\omega_f$).
%\widetext
\begin{widetext}
The phase is then given by
\begin{eqnarray}
\Psi  &=& {1 \over 32\eta} \Biggl\{ \left[ (m\omega_i)^{-5/3}
- (m\omega_f)^{-5/3} \right]
+ {5 \over 1008}(743+924\eta) \left[ (m\omega_i)^{-1} - (m\omega_f)^{-1}
\right]
\nonumber \\ && \mbox{}
+ \left[ {5 \over 24} \sum_{i = 1,2} \left[ \chi_i ({\bf \hat L_N
\cdot \hat s_i}) ( 113{m_i^2
\over m^2} +75\eta) \right] - 10\pi \right] \left[
(m\omega_i)^{-2/3} - (m\omega_f)^{-2/3} \right]  \nonumber \\ && \mbox{}
+ {5 \over 48}
\eta \chi_1 \chi_2 \left[ 247({\bf \hat s_1 \cdot \hat s_2})
- 721({\bf \hat L_N
\cdot \hat s_1})({\bf \hat L_N \cdot \hat s_2}) \right] \left[
(m\omega_i)^{-1/3} - (m\omega_f)^{-1/3} \right] \Biggr\}. \label{phase}
\end{eqnarray}
The spin dependence gives the acceleration or delay of the spin orbit
coupling, while it is also crucial to account for the change with spin 
of the final frequency $\omega_f$. 
%Note that in the particle limit, we can take $\omega_f\sim\omega_{isco}$. \cite{Bardeen:1972fi} \cite{Favata:2010ic}

\end{widetext}

In our notation, the leading PN-dependence is given by
\begin{equation}
\label{eq:SPN}
\frac{188}{113}S_{PN}=\left(\vec{S}\cdot\hat{L}+\frac{75}{188}\,\delta{m}\vec{\Delta}\cdot\hat{L}\right),
\end{equation}
with $75/188=0.3989$.

\section{Simulations}\label{sec:Simulations}
%Still rough version to be revisited

In order to evaluate the hangup effect dependence on the spins and
the mass ratio of the binary we will make use of the 107 simulations
we selected from the Refs. \cite{Healy:2014yta,Healy:2016lce,Healy:2017psd}
and the current 74 presented
in this paper. In order to quantify this hangup effect we count 
the number of orbits to merger (as measured by the peak of the amplitude
of the (2,2)-mode of the $h$ waveform) from an initial 
fiducial orbital frequency of $\omega_{i}=0.07$. This value of
$\omega_{i}$ is chosen such that all the 181 simulations include 
cleanly this and higher frequencies in their waveforms
The number of orbits are computed in an invariant way (as opposed to
coordinate tracks) by counting (half) the number of cycles of the 
(2,2)-mode waveforms (extrapolated to an infinite observer location 
via \cite{Nakano:2015pta}).
Table \ref{tab:ecc} provides an account of the relevant parameters in this
regard for the new 74 simulations.

The spatial resolution of each simulation can be described by a number NXXX,
where XXX is related to the resolution of the grid in the wavezone.  For example,
a resolution tag of N140 would have resolution of $M/1.40$ in the wavezone.
This global resolution factor is chosen such that the mass and spin are conserved
to an acceptable degree, and in accordance with the convergence studies conducted
in Refs.\cite{Healy:2014yta,Healy:2016lce}.
The new runs presented here are in 3 families: $q=0.85$ with 
resolution N120, $q=0.4142$ with resolution N100, and $q=0.20$ with resolution
N120.  From each family, a sample of simulations are produced at 3 resolutions
to verify accuracy. Other additional runs added not in these series have 
resolutions of N100, N120, or N140.

We will study the hangup dependence of those 181 simulations on the variable
\beq
\frac{1}{1-C}S_{hu}=\left(\vec{S}\cdot\hat{L}+C\,\delta{m}\vec{\Delta}\cdot\hat{L}\right),
\eeq
where $C$ will be the fitting parameter that regulates the coupling
to the total spin $\vec{S}$ with the ``delta'' combination 
$\delta{m}\vec{\Delta}$.

Note that our study does not need to make reference to post-Newtonian
expansions and uses only full numerical evolutions. The above variables
in common with PN can be independently obtained from symmetry considerations
(parity and exchange of $1\longleftrightarrow2$ BH labels) as discussed in 
\cite{Lousto:2012gt,Lousto:2013wta}.

\begin{figure}
  \includegraphics[angle=270,width=0.99\columnwidth]{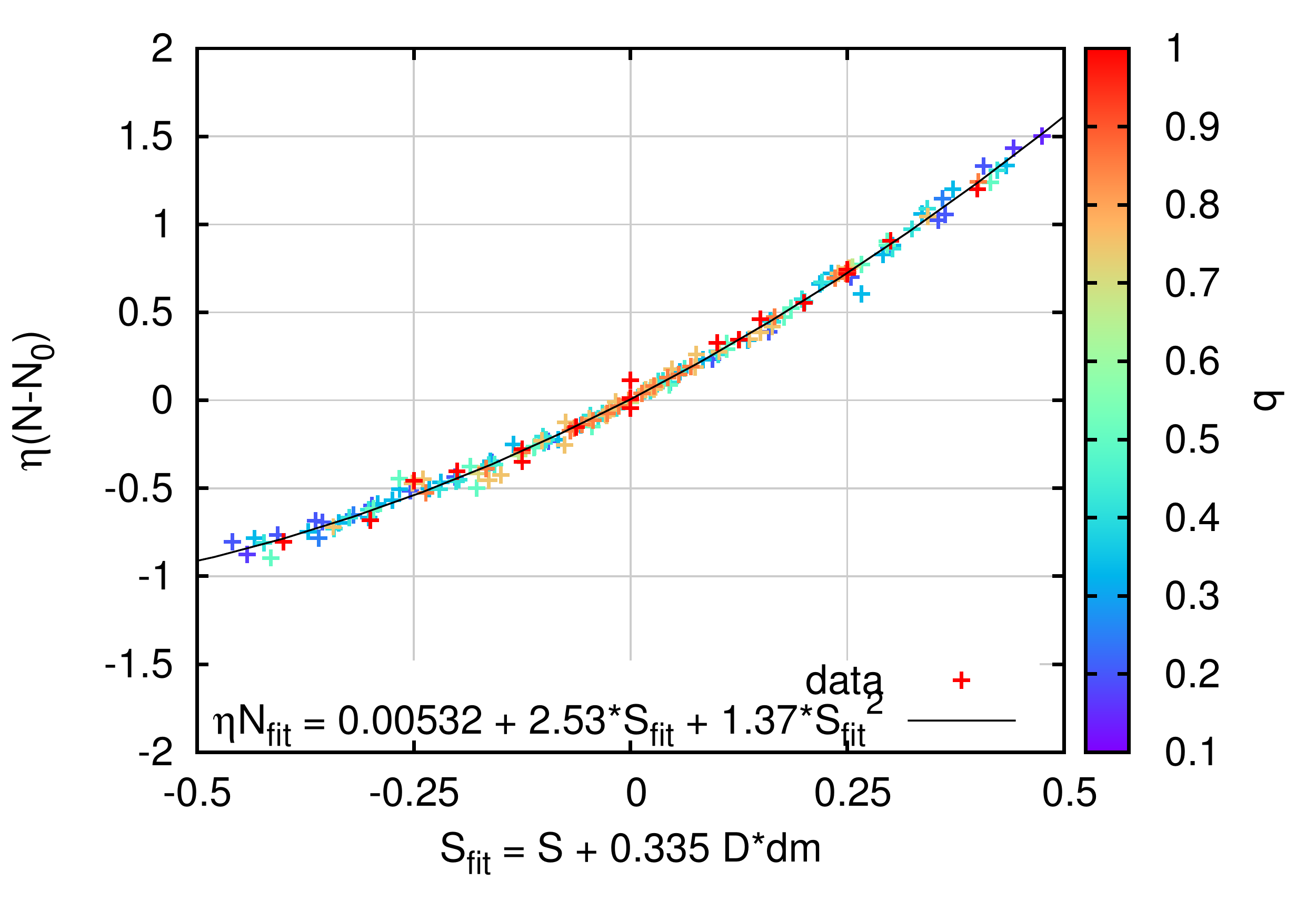}\\
  \includegraphics[angle=270,width=0.99\columnwidth]{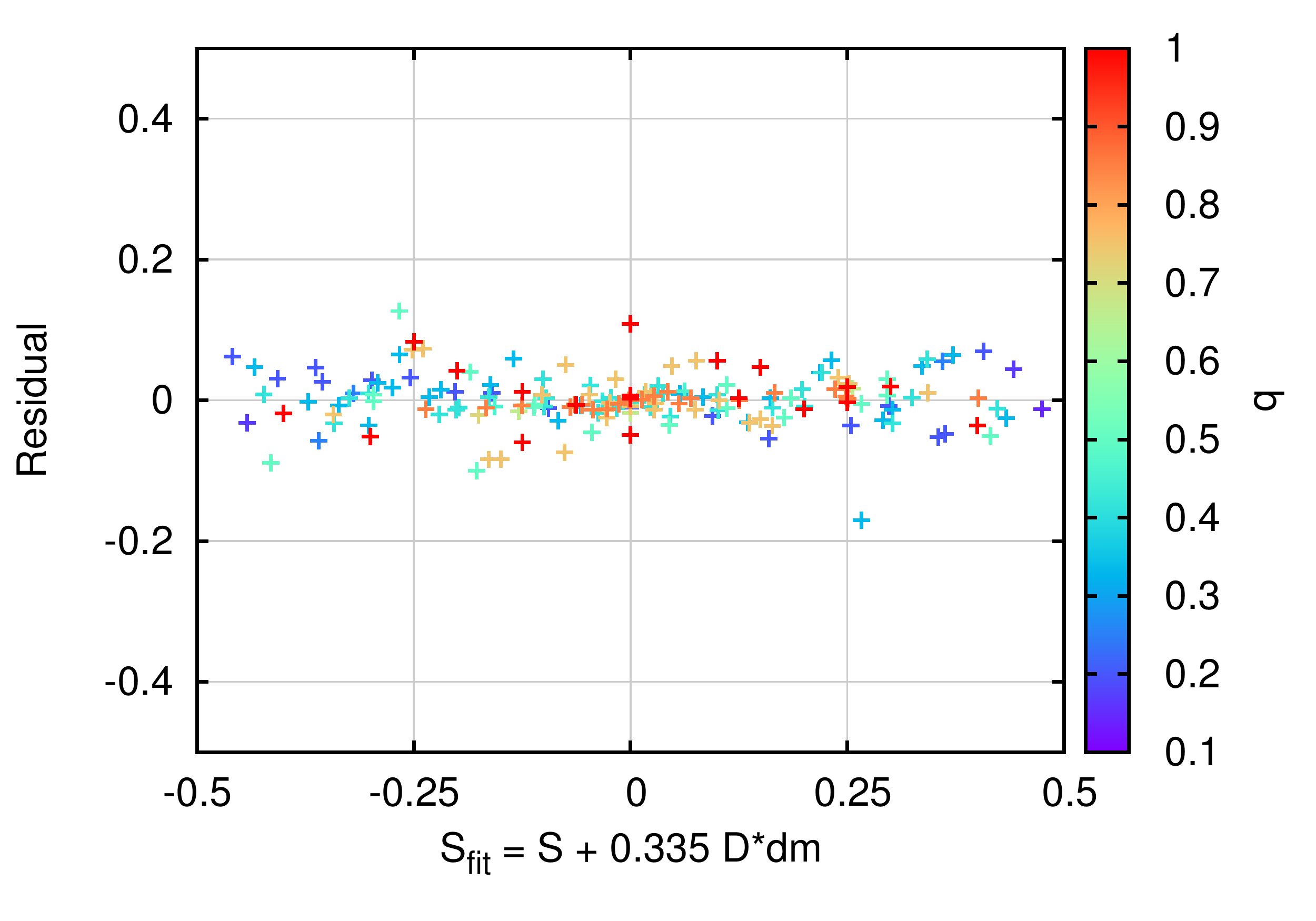}
  \caption{The number of orbits differential with respect to the nonspinning
case for full numerical binary black hole mergers. We use the (2,2) mode 
of the waveform and calculate the number of cycles between $m\omega=0.07$ and
$m\omega_{peak}$. We studied in detail the cases with
$q=1.00$, $q=0.85$, $q=0.75$, $q=0.4142$, $q=0.50$, $q=0.333$ and $q=0.20$ 
and fit a quadratic dependence with the spin variables to extract the
linear spin coefficients of $\vec{S}\cdot\hat{L}+C \delta{m}\vec{\Delta}\cdot\hat{L}$. The residuals of such fit are also displayed showing no systematics.
\label{fig:hangupNR}}
\end{figure}

The results of a fitting of the form 
(where $N$ is the number of orbits to merger for spinning binaries and
$N_0$ the corresponding for nonspinning binaries from the same initial
fiducial orbital frequency)
\beq\label{eq:Shufit}
\eta[N-N_0]=D+A\,S_{hu}+B\,S_{hu}^2,
\eeq
are presented in
Fig.~\ref{fig:hangupNR}. This shows the dependence of the hangup
effect with respect to the nonspinning binaries. We see
that this dependence can be expressed in terms of the spin variable
\beq\label{eq:Shu}
\frac32S_{hu}=\left(\vec{S}\cdot\hat{L}+\frac{1}{3}\,\delta{m}\vec{\Delta}\cdot\hat{L}\right),
\eeq
to an excellent degree of approximation since $C=0.3347$ from the fits.

Note the small residual coefficient, 0.00532, for vanishing spins 
displaying the consistent subtraction of the nonspinning portion 
even for spinning binaries. The residuals panel on the bottom of
Fig.~\ref{fig:hangupNR} shows that all residuals are an order of magnitude
smaller than its fit range above.

Table \ref{tab:RMS} displays the comparative statistical properties of the
fits if we use the alternative variables $S_0$ or $S_{eff}$ as given
in Eq.~(\ref{eq:S0Seff}) and $S_{PN}$ as given in Eq.~(\ref{eq:SPN}.

\begin{table}
\caption{RMS and variance of $S_0$, $S_{eff}$, and $S_{hu}$ fits.
ndf (no. degrees of freedom), WSSR = weighted sum of the residuals
RMS=$\sqrt{WSSR/ndf}$, Variance=reduced $\chi^2$ = WSSR/ndf}
\label{tab:RMS}
\begin{ruledtabular}
\begin{tabular}{cccccc}
Variable & Coefficient & ndf & WSSR & RMS & Variance\\
\hline
$S_0$     & 0.5      & 167 & 0.702 & 0.065 & 0.0042\\
$S_{eff}$ & 0.428571 & 167 & 0.361 & 0.047 & 0.0022 \\
$S_{PN}$  & 0.398936 & 167 & 0.281 & 0.041 & 0.0017 \\
$S_{hu}$  & 0.333333 & 167 & 0.214 & 0.036 & 0.0013
\end{tabular}
\end{ruledtabular}
\end{table}

%%%%%%%%%%%

As a control study of the above results we designed two sequences
of runs that check if there is a null hangup effect when either
$\vec{S}=0$ or $\vec{S}_0=0$.

By requiring that $\vec{S}=0$ we get
\begin{equation}
\vec\alpha_2=-q^2\vec\alpha_1
\end{equation}
hence
\begin{equation}
\delta{m}\,\tilde{\vec\Delta}(\vec{S}=0)=\frac{(1-q)\,q\,\vec{\alpha}_1}{(1+q)}
\end{equation}

The maximum effect hence occurs for $q^{max}=\sqrt{2}-1$ for any magnitude
of $\alpha_1$.

We choose a few representative cases for $\alpha_1=0,\pm0.4,\pm0.8$ to
model the effect. that lead to $\alpha_2=0,\mp0.0686,\mp0.13726$.

If we want to compare to something that we suspect will be closer to mimic the nonspinning case, we can set $S_0=0$. In that case
\begin{equation}\label{eq:S0}
\tilde{\vec{S}}_0=(\vec\alpha_2+q\vec\alpha_1)/(1+q)=0,
\end{equation}
implies
\begin{equation}
\vec\alpha_2=-q\vec\alpha_1
\end{equation}
which gives 
\begin{equation}
\tilde{\vec{S}}(\vec{S}_0=0)=-\frac{(1-q)\,q\,\vec{\alpha}_1}{(1+q)^2}
\end{equation}
with a $q^{max}=1/3$.

One can check that 
\begin{equation}
\delta{m}\,\tilde{\vec\Delta}(\vec{S}_0=0)=-2\tilde{\vec{S}}(\vec{S}_0=0).
\end{equation}

Since for $q=1/3$ or $q=\sqrt{2}-1$, $S(q)$ does not change by much around the maximum (-1/8 vs. -0.121), we can still use $q=\sqrt{2}-1$ as the reference $q$,
with the advantage of direct comparison for the case $S=0$. Hence
we study 4 new runs with $\alpha_1=\pm0.4,\pm0.8$ to
model the effect. that lead to $\alpha_2=\mp0.1657,\mp0.33137$.

A parameter space view of the runs we performed for those families (and others)
is in the of Fig.~\ref{fig:runs} labeled as $q=0.4142$.

%%%%%%%%%%%

\begin{figure}
\includegraphics[angle=270,width=0.85\columnwidth]{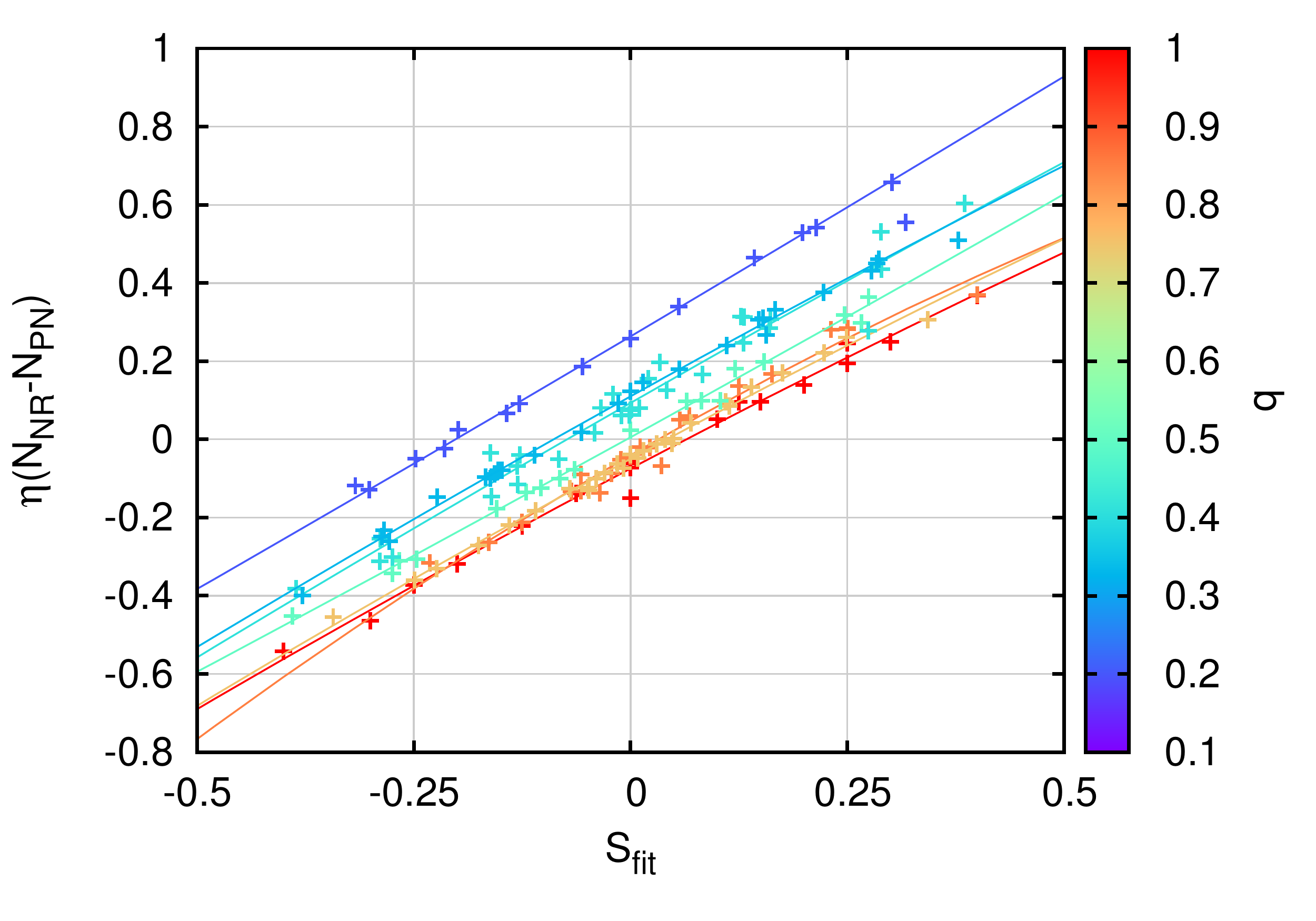}\\
\includegraphics[angle=270,width=0.85\columnwidth]{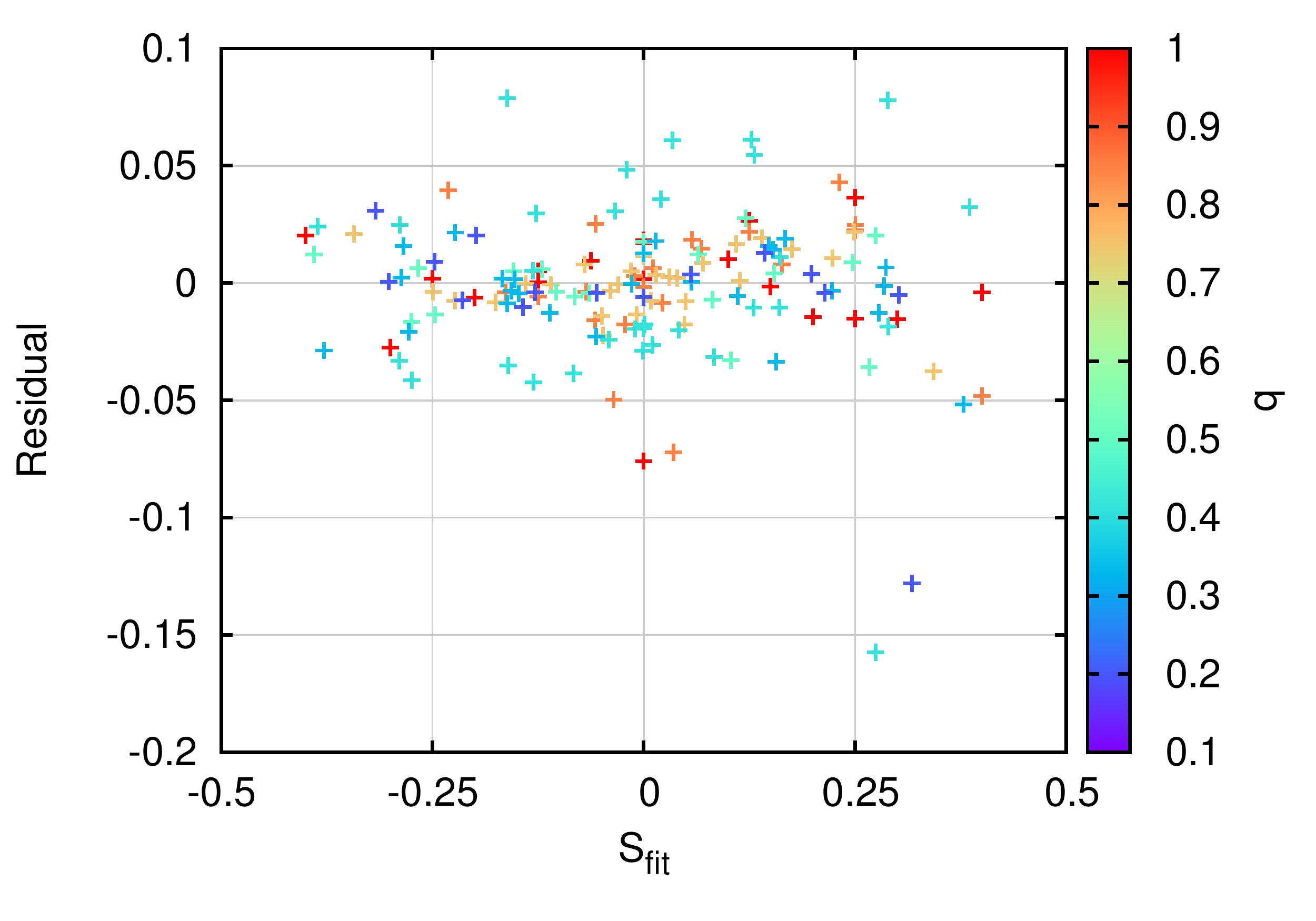}
\caption{Differential hangup effect NR vs. PN of the simulated binaries for 
$q=1.00$, $q=0.85$, $q=0.75$,  $q=0.50$, $q=0.4142$, $q=0.333$, and $q=0.2$ displaying the stronger dependence
on spins in full numerical simulations than predicted by PN and
the spin variable deviation from simply $\vec{S}\cdot\hat{L}$. An additional
dependence with $q$ is also observed.  
\label{fig:hangupPN}}
\end{figure}

Another control study is to try to perform similar studies with purely
3.5PN evolutions as used in Ref.~\cite{Lousto:2015uwa}.
Figure~\ref{fig:hangupPN} displays a measure of the 
differential hangup effect
further delaying or prompting merger of the full numerical
evolutions with respect to the 3.5PN
integrations. This residual differences (also depending on $q$)
gives us a measure of how much stronger the
effect is in full General Relativity. We also see that the 
variable that describes the effect is not simply $\vec{S}\cdot\vec{L}$,
but rather $S_{fit}$ something proportional to 
$(\vec{S}+(0.53\pm0.08)\delta{m}\vec{\Delta})\cdot\vec{L}$ (see Table \ref{tab:hangupPN_fits}), 
that do not corresponds
to the $\vec{S}_0\cdot\vec{L}$ variable, that is a quasi-conserved
quantity~\cite{Racine:2008qv}.

\begin{table}
\caption{Table of fitting coefficients for each line in Fig.~\ref{fig:hangupPN}. The fit is of the 
form $\eta(N_{NR}-N_{PN}) = D + AS_{fit} + BS_{fit}^2$ where $S_{fit} = S + C \delta m \Delta$.
}
\label{tab:hangupPN_fits}
\begin{ruledtabular}
\begin{tabular}{ccccc}
$q$ & $A$ & $B$ & $C$ & $D$ \\
\hline
1.00 & 1.167	&	-0.092	&	0	&	-0.075 \\
0.85 & 1.281	&	-0.207	&	0.453	&	-0.041  \\ 
0.75 & 1.194	&	-0.097	&	0.580	&	-0.050 \\
0.50 & 1.222	&	 0.029	&	0.612	&	 0.006 \\
0.4142 & 1.266	&	-0.044	&	0.605	&	 0.093 \\
0.3333 & 1.231	&	-0.068	&	0.608	&	 0.110 \\
0.2  & 1.311	&	 0.022	&	0.536	&	 0.263 \\
\end{tabular}
\end{ruledtabular}
\end{table}

We conclude that there are residual effects at PN level that are not as simply
parameterized as for the purely full numerical evolutions with $S_{hu}$.

\section{Peak luminosity, amplitude and frequency modeling}\label{sec:Omegapeak}

The end of the inspiral of two black holes is characterized by a plunge towards the formation of a highly distorted final single black hole. It is during this process that the black holes radiates the most power in the form of gravitational waves. One can thus identify the peak luminosity and the corresponding
amplitude and the frequency (derived from the phase) of the gravitational waveforms. These quantities are of interest for gravitational wave observations 
%\cite{TheLIGOScientific:2016wfe,TheLIGOScientific:2016pea,Keitel:2016krm}
and could be used as potential tests of general relativity
(if measured independently) as there are theory of gravity 
specific relationships among them (as mentioned in \cite{Healy:2017mvh}).
Other test of general relativity have been described and applied to
observations in
\cite{TheLIGOScientific:2016src,TheLIGOScientific:2016pea,Abbott:2017vtc}.

In this section we make use of the new set of simulations to provide 
a more accurate modeling of the peak luminosity, amplitude and frequency.

\subsection{Peak luminosity modeling}\label{sec:Plum}

In Ref.~\cite{Healy:2016lce} we proposed the following fourth order expansion to fit the peak luminosity

\begin{eqnarray}\label{eq:4plum}
L_{\rm peak} = (4\eta)^2\,\Big\{N_0 + N_1 \Spar+ N_{2a}\,\Dpar\dmt+\nonumber\\
                     N_{2b}\,\Spar^2+ 
                     N_{2c}\,\Dpar^2+
                     N_{2d}\,\dmt^2 +\nonumber\\
                     N_{3a}\,\Dpar\Spar\dmt+ 
                     N_{3b}\,\Spar\Dpar^2+
                     N_{3c}\,\Spar^3+ \nonumber\\
                     N_{3d}\,\Spar\dmt^2+
                     N_{4a}\,\Dpar\Spar^2\dmt + \nonumber\\
                     N_{4b}\,\Dpar^3\dmt +
                     N_{4c}\,\Dpar^4+
                     N_{4d}\,\Spar^4+\nonumber\\
                     N_{4e}\,\Dpar^2 \Spar^2+
                     N_{4f}\,\dmt^4+
                     N_{4g}\,\Dpar\dmt^3+\nonumber\\
                     N_{4h}\,\Dpar^2\dmt^2 +
                     N_{4i}\,\Spar^2\dmt^2\Big\}.
\end{eqnarray}
Where all $N_i$ are fitting parameters (as used in  Ref.~\cite{Healy:2014yta}).

\begin{figure}
\includegraphics[angle=270,width=0.99\columnwidth]{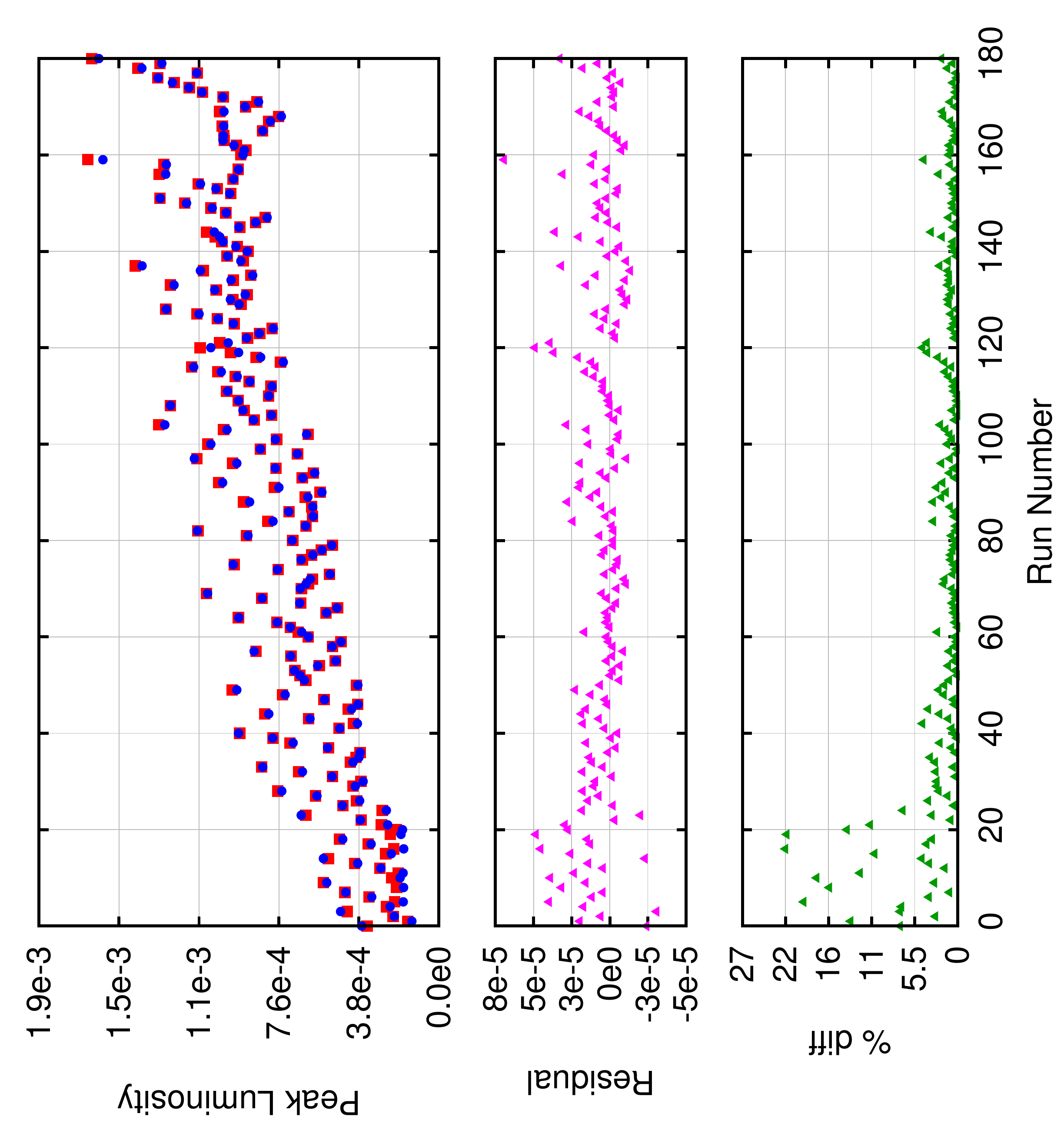}
\caption{Top panel: Red squares representing data and blue dots their
  corresponding fit to the peak luminosity. Bottom panels:
  Fitting residuals of the peak luminosity formula as given in
  Eq.~(\ref{eq:4plum}) (See also  \cite{Healy:2016lce}).
\label{fig:pLum}}
\end{figure}

In Fig.~\ref{fig:pLum} we display the agreement between the
peak luminosity formula Eq.~(\ref{eq:4plum})
(See also Ref.~\cite{Healy:2016lce}) with the whole set
of 181 simulations provided in this paper.
We have fitted to 16 out of the 19 coefficients here by choosing
the 3 spinless coefficients, $N_0, N_{2d}$, and $N_{4f}$ to match
the values found in Ref.~\cite{Healy:2017mvh}
after an exceptionally accurate convergence study.
Fitting independently all 19 coefficients produce values close 
to those assumed for the 3 nonspinning ones.
This leads to larger residuals for the spin dependence, but
an overall higher accuracy of the fitting formula. This
hierarchical approach is similar to assigning the highest
weight to those extrapolated nonspinning 
waveforms in ~\cite{Healy:2017mvh}.
The explicit values of those parameters as well as those
obtained in the fitting here are provided in the appendix
Table \ref{tab:fitparsVL}.

\subsection{Peak amplitude and frequency modeling}\label{sec:PAOm}

In Ref.~\cite{Healy:2017mvh} we modeled the peak amplitude
and peak frequency for the nonspinning binaries
(See also independent studies in
\cite{Pan:2011gk,Taracchini:2012ig,Bohe:2016gbl}).
Here we generalize those fitting formulae for the aligned spinning binary black hole mergers.

\begin{eqnarray}\label{eq:PAmp}
(r/m)h_{22}^{\rm peak} = (4\eta)\,\Big\{H_0 + H_1 \Spar+ H_{2a}\,\Dpar\dmt+\nonumber\\
                     H_{2b}\,\Spar^2+
                     H_{2c}\,\Dpar^2+
                     H_{2d}\,\dmt^2 + \nonumber\\
                     H_{3a}\,\Dpar\Spar\dmt+
                     H_{3b}\,\Spar\Dpar^2+
                     H_{3c}\,\Spar^3+ \nonumber\\
                     H_{3d}\,\Spar\dmt^2+
                     H_{4a}\,\Dpar\Spar^2\dmt + \nonumber\\
                     H_{4b}\,\Dpar^3\dmt +
                     H_{4c}\,\Dpar^4+
                     H_{4d}\,\Spar^4+\nonumber\\
                     H_{4e}\,\Dpar^2 \Spar^2+
                     H_{4f}\,\dmt^4+
                     H_{4g}\,\Dpar\dmt^3+\nonumber\\
                     H_{4h}\,\Dpar^2\dmt^2 +
                     H_{4i}\,\Spar^2\dmt^2\Big\}.
\end{eqnarray}
With all $H_i$ fitting parameters.

\begin{figure}
\includegraphics[angle=270,width=0.99\columnwidth]{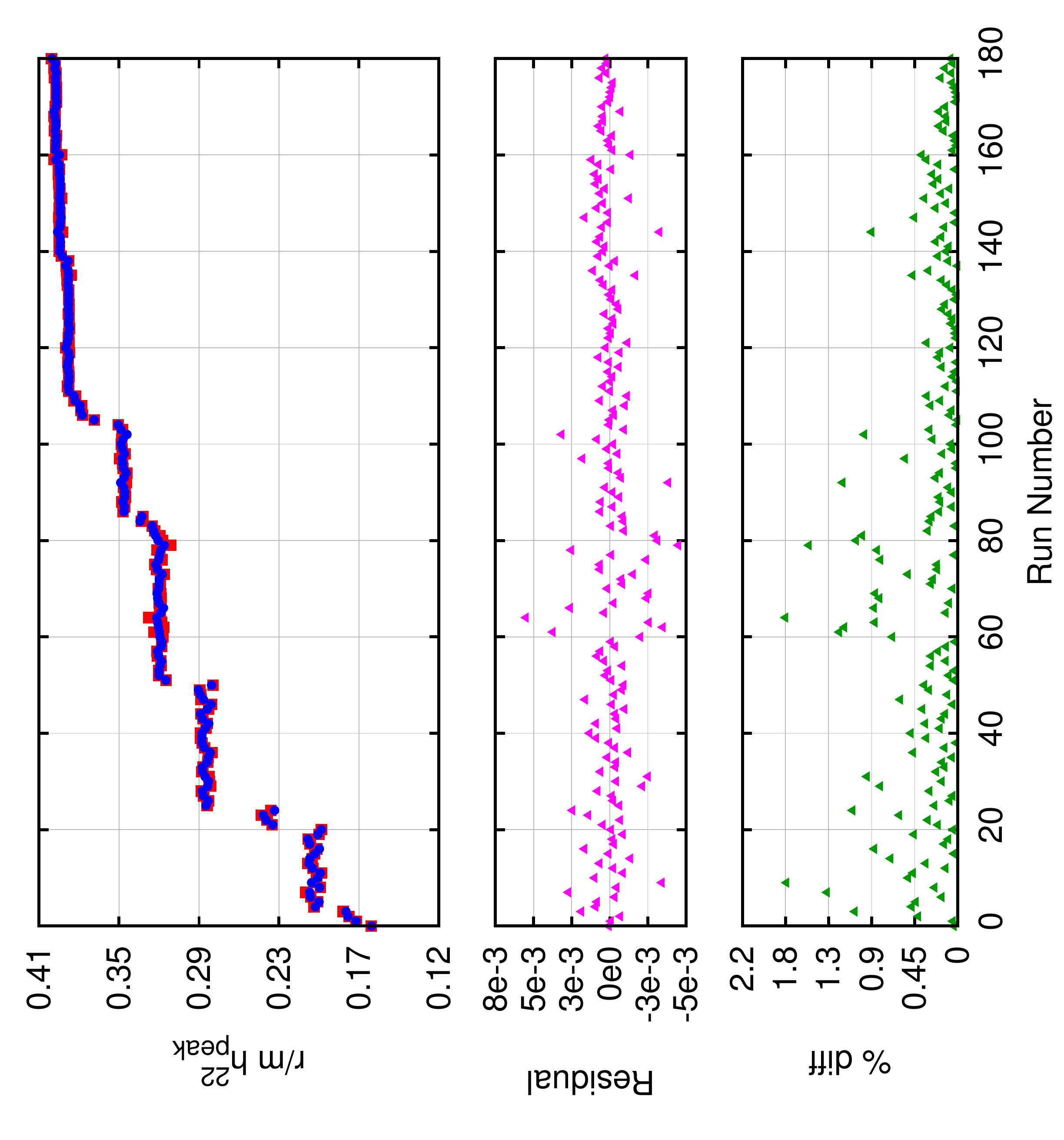}
\caption{Top panel: Red squares representing data and blue dots their
  corresponding fit to the peak amplitude of the (2,2)-mode. Bottom panels:
  Fitting residuals of the peak amplitude formula as given in
  Eq.~(\ref{eq:PAmp}).
\label{fig:PAmp}}
\end{figure}

In Fig.~\ref{fig:PAmp} we display the agreement between the
new peak amplitude formula given here with the updated set
of simulations provided in this paper.
We have fitted to 16 out of the 19 coefficients here by choosing
the 3 spinless coefficients, $H_0, H_{2d}$, and $H_{4f}$ to match
the values found in Ref.~\cite{Healy:2017mvh}
after an extrapolation of accurate convergence sequence.
This choice, while producing slightly larger residuals, should
produce a more accurate overall fit.
The explicit values of those parameters as well as those
obtained in the fitting here are provided in the appendix
Table \ref{tab:fitparsOA}.

Following again the introduction of the peak frequency for spinless binaries in
 Ref.~\cite{Healy:2017mvh} we generalize those
fitting formulae for the aligned spinning binary black hole mergers
\begin{eqnarray}\label{eq:Pfreq}
m\omega_{22}^{\rm peak} = \Big\{W_0 + W_1 \Spar+ W_{2a}\,\Dpar\dmt+\nonumber\\
                     W_{2b}\,\Spar^2+ 
                     W_{2c}\,\Dpar^2+
                     W_{2d}\,\dmt^2 +\nonumber\\
                     W_{3a}\,\Dpar\Spar\dmt+ 
                     W_{3b}\,\Spar\Dpar^2+
                     W_{3c}\,\Spar^3+ \nonumber\\
                     W_{3d}\,\Spar\dmt^2+
                     W_{4a}\,\Dpar\Spar^2\dmt + \nonumber\\
                     W_{4b}\,\Dpar^3\dmt +
                     W_{4c}\,\Dpar^4+
                     W_{4d}\,\Spar^4+\nonumber\\
                     W_{4e}\,\Dpar^2 \Spar^2+
                     W_{4f}\,\dmt^4+
                     W_{4g}\,\Dpar\dmt^3+\nonumber\\
                     W_{4h}\,\Dpar^2\dmt^2 +
                     W_{4i}\,\Spar^2\dmt^2\Big\}.
\end{eqnarray}
With all $W_i$ fitting parameters.

\begin{figure}
\includegraphics[angle=270,width=0.99\columnwidth]{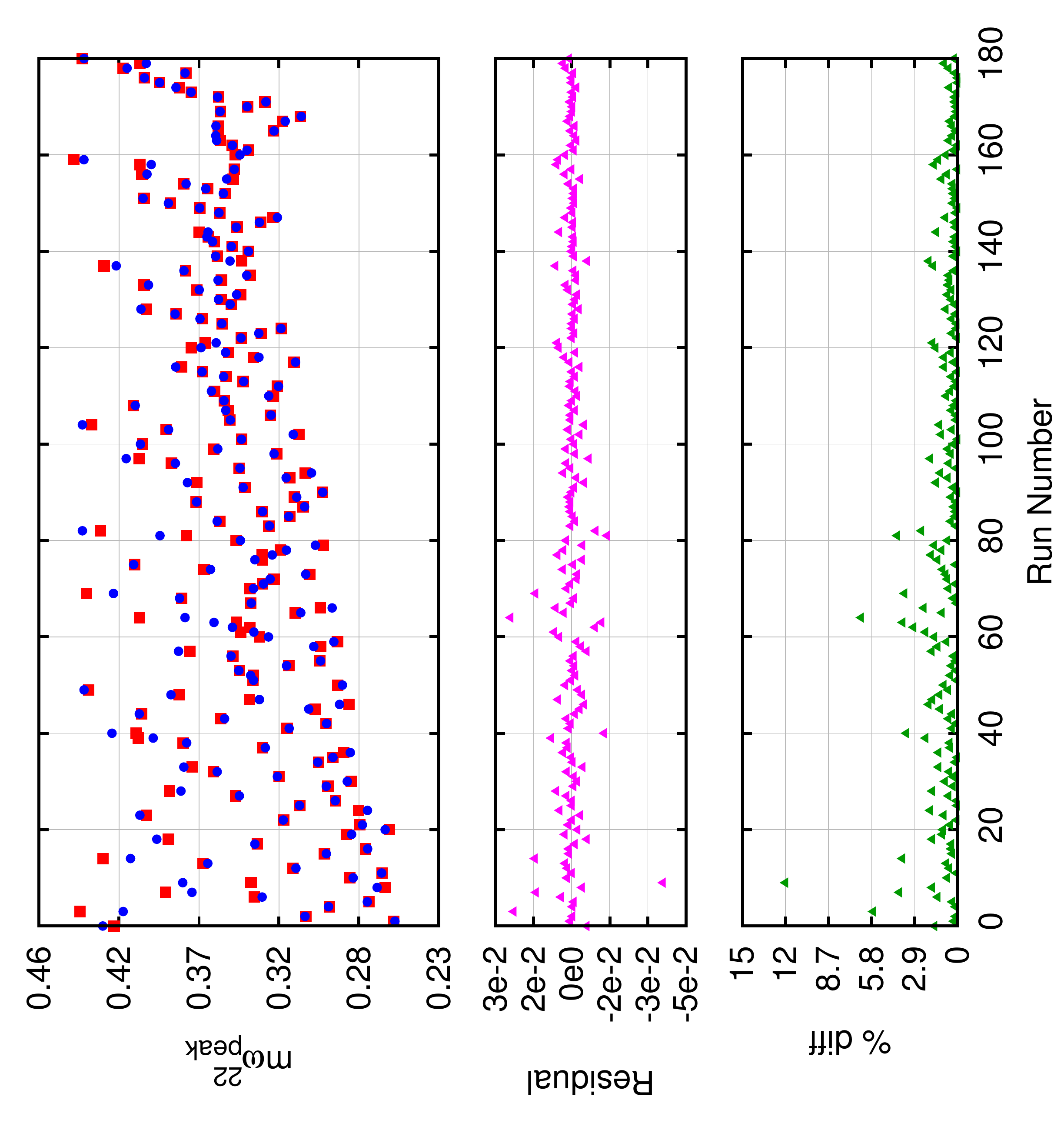}
\caption{Top panel: Red squares representing data and blue dots their
  corresponding fit to the frequency of the peak amplitude of the
(2,2)-mode. Bottom panels:
  Fitting residuals of the frequency at peak amplitude formula as given in
  Eq.~(\ref{eq:Pfreq}).
\label{fig:OAmp}}
\end{figure}

In Fig.~\ref{fig:OAmp} we display the agreement between the
new peak amplitude formula given here with the updated set
of 181 simulations provided in this paper.
We have fitted to 16 out of the 19 coefficients here by choosing
the 3 spinless coefficients, $W_0, W_{2d}$, and $W_{4f}$ to match
the values found in Ref.~\cite{Healy:2017mvh}
after an extrapolation of an accurate three convergence sequence.
Matching all 19 coefficients leads to values close to those
previous work~\cite{Healy:2017mvh} for the nonspinning case,
hence we assumed those values for our reduced fits. 
The explicit values of those parameters as well as those
obtained in the fitting here are provided in the appendix \ref{app:fit},
Table \ref{tab:fitparsOA}.

In summary, we find the fitting statistics as given in Table~\ref{tab:Newstats}

\begin{table}
\caption{Fitting statistics for peak luminosity, frequency and amplitude of the mode (2,2) formulae
}
\label{tab:Newstats}
\begin{ruledtabular}
\begin{tabular}{cccc}
Value &	Peak Luminosity	& Peak $m\omega_{22}$ &	Peak $(r/m) h_{22}$\\
\hline
RMS 	&	1.68809e-05&	5.95755e-03&	1.54105e-03\\
Std. Dev.&      1.46664e-05&    5.70386e-03&    1.47523e-03\\
Avg. Diff.&	6.77198e-06&	2.70026e-05&	-2.44938e-05\\
Max Diff.&	7.18966e-05&	2.63070e-02&	8.39437e-03\\
Min Diff.&	-3.18964e-05&	-3.94535e-02&	-4.81573e-03
\end{tabular}
\end{ruledtabular}
\end{table}

We expect that the hierarchical approach followed in these fitting
(use of accurate 3-parameters from the nonspinning binary cases),
provides an accurate account of the phenomenology of the peak
emission of gravitational waves.
See also, for instance, the approach in
Ref.~\cite{Jimenez-Forteza:2016oae}.

\section{Remnant modeling}\label{sec:Remnant}

The modeling of the final mass and spin as well as the recoil of the final
merged black hole has been the subject of many studies ever since the 
numerical relativity breakthroughs 
\cite{Pretorius:2005gq,Campanelli:2005dd,Baker:2005vv}
allowed the long term evolutions of
binary black holes. The interest for such formulae have been recently
renewed
\cite{Jimenez-Forteza:2016oae,Ghosh:2016qgn,Ghosh:2017gfp}
as they provide important information for the modeling of 
waveforms \cite{Khan:2015jqa,Bohe:2016gbl}
and interpretation of the gravitational wave observations
as well as providing consistency test for general relativity
\cite{TheLIGOScientific:2016src,Abbott:2017vtc}
Below we make use to improve (notably for the recoil velocity)
the current fitting formulae for the
remnant properties of the final black hole with the new set of simulations.

\subsection{Final Mass modeling}\label{sec:Mass}

In Ref.~\cite{Healy:2014yta} the fitting formula for the remnant mass
$M_{\rm rem}$ was given by,

\begin{eqnarray}\label{eq:4mass}
\frac{M_{\rm rem}}{m} = (4\eta)^2\,\Big\{M_0 + K_1 \Spar+ K_{2a}\,\Dpar\dmt +\nonumber\\
                     K_{2b}\,\Spar^2+ 
                     K_{2c}\,\Dpar^2+
                     K_{2d}\,\dmt^2 +\nonumber\\
                     K_{3a}\,\Dpar\Spar\dmt+ 
                     K_{3b}\,\Spar\Dpar^2+
                     K_{3c}\,\Spar^3+ \nonumber\\
                     K_{3d}\,\Spar\dmt^2+
                     K_{4a}\,\Dpar\Spar^2\dmt + \nonumber\\
                     K_{4b}\,\Dpar^3\dmt +
                     K_{4c}\,\Dpar^4+
                     K_{4d}\,\Spar^4+\nonumber\\
                     K_{4e}\,\Dpar^2 \Spar^2+
                     K_{4f}\,\dmt^4+
                     K_{4g}\,\Dpar\dmt^3+\nonumber\\
                     K_{4h}\,\Dpar^2\dmt^2 +
                     K_{4i}\,\Spar^2\dmt^2\Big\}+\nonumber\\
                     \left[1+\eta(\tilde{E}_{\rm ISCO}+11)\right]\dmt^6 .
\end{eqnarray}
With all 19 $K_i$ being fitting parameters.

\begin{figure}
\includegraphics[angle=270,width=0.99\columnwidth]{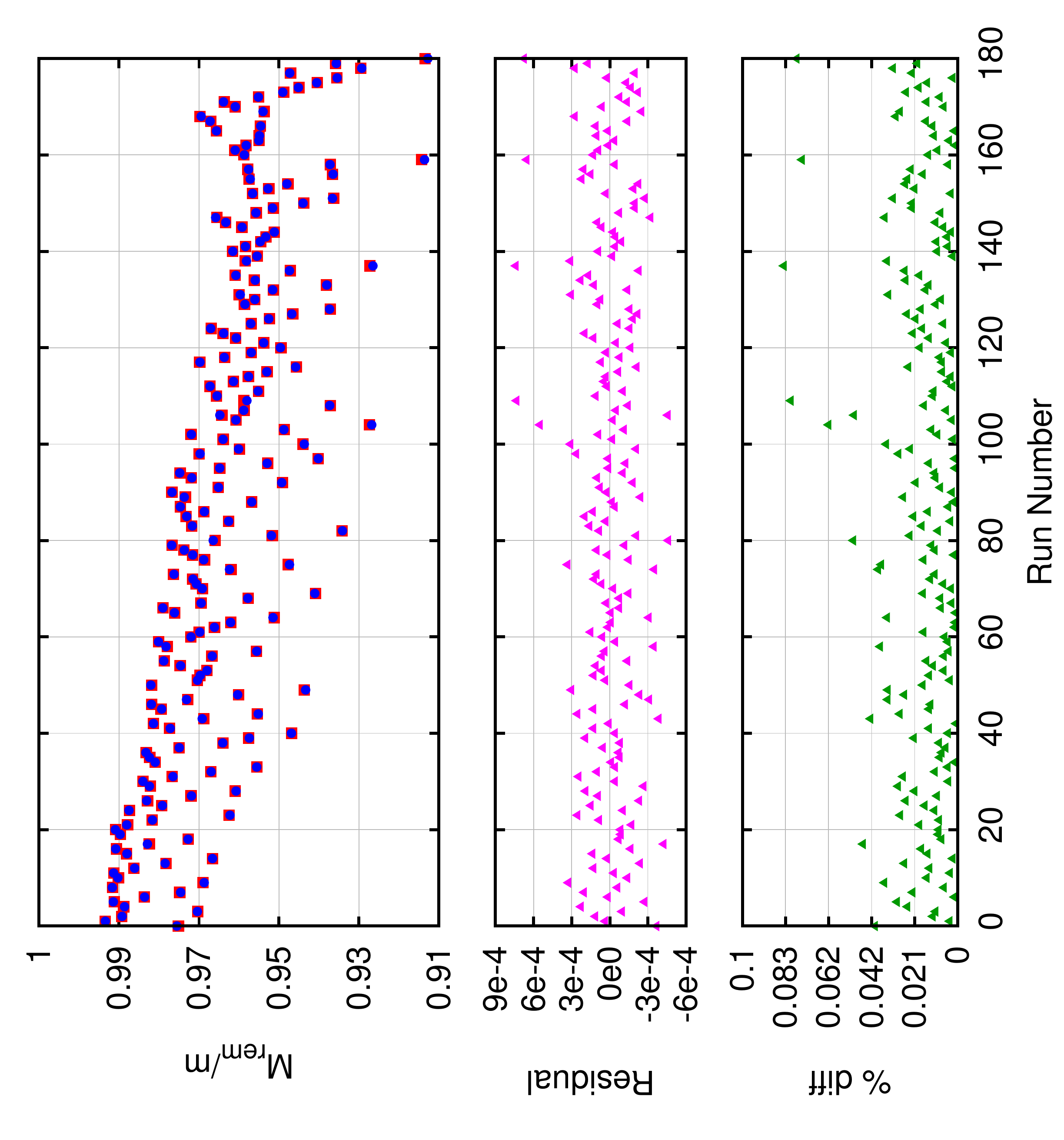}
\caption{Top panel: Red squares representing data and blue dots their
  corresponding fit. Bottom panels:
  Fitting residuals of the final remnant mass formula as given in
Eq.~(\ref{eq:4mass})  (See also Ref.~\cite{Healy:2016lce}).
\label{fig:mass}}
\end{figure}

In Fig.~\ref{fig:mass} we display the agreement between the
latest final mass formula as in Eq.~(\ref{eq:4mass})
(See also Ref.~\cite{Healy:2016lce}) with the whole set
of simulations provided in this paper.
Table \ref{tab:fitparsms} gives the 19 parameters for the final mass
optimal fit.
Making use of the accurate determination of the final mass
via the isolated horizon formalism~\cite{Dreyer02a} we observe
that those coefficients with nearly vanishing values can be
adopted as precisely zero. Thus, Table \ref{tab:fitparsms_reduced} 
gives an alternative reduced set of 9 parameters fit.
This may provide a helpful approach to extend these
formulae to the precessing binaries case.

\subsection{Final Spin Modeling}\label{sec:Spin}

As in Ref.~\cite{Healy:2014yta} the fitting formula for the final spin has the form,

\begin{eqnarray}\label{eq:4spin}
\alpha_{\rm rem} = \frac{S_{\rm rem}}{M^2_{\rm rem}} =
                     (4\eta)^2\Big\{L_0 + L_{1}\,\Spar+\nonumber\\ 
                     L_{2a}\,\Dpar\dmt+
                     L_{2b}\,\Spar^2+
                     L_{2c}\,\Dpar^2+
                     L_{2d}\,\dmt^2+\nonumber\\
                     L_{3a}\,\Dpar\Spar\dmt+
                     L_{3b}\,\Spar\Dpar^2+
                     L_{3c}\,\Spar^3+\nonumber\\
                     L_{3d}\,\Spar\dmt^2+
                     L_{4a}\,\Dpar\Spar^2\dmt+
                     L_{4b}\,\Dpar^3\dmt+\nonumber\\
                     L_{4c}\,\Dpar^4+
                     L_{4d}\,\Spar^4+
                     L_{4e}\,\Dpar^2\Spar^2+\nonumber\\
                     L_{4f}\,\dmt^4+
                     L_{4g}\,\Dpar\dmt^3+\nonumber\\
                     L_{4h}\,\Dpar^2\dmt^2+
                     L_{4i}\,\Spar^2\dmt^2\Big\}+\nonumber\\
                     \Spar(1+8\eta)\dmt^4+\eta\tilde{J}_{\rm ISCO}\dmt^6.
\end{eqnarray}
With 19 $L_i$ fitting parameters.

Note that the two formulae, for the remnant mass and spin, above impose the particle limit by including
the ISCO dependencies (See Ref. \cite{Healy:2014yta,Ori:2000zn} for the explicit expressions).

\begin{figure}
\includegraphics[angle=270,width=0.99\columnwidth]{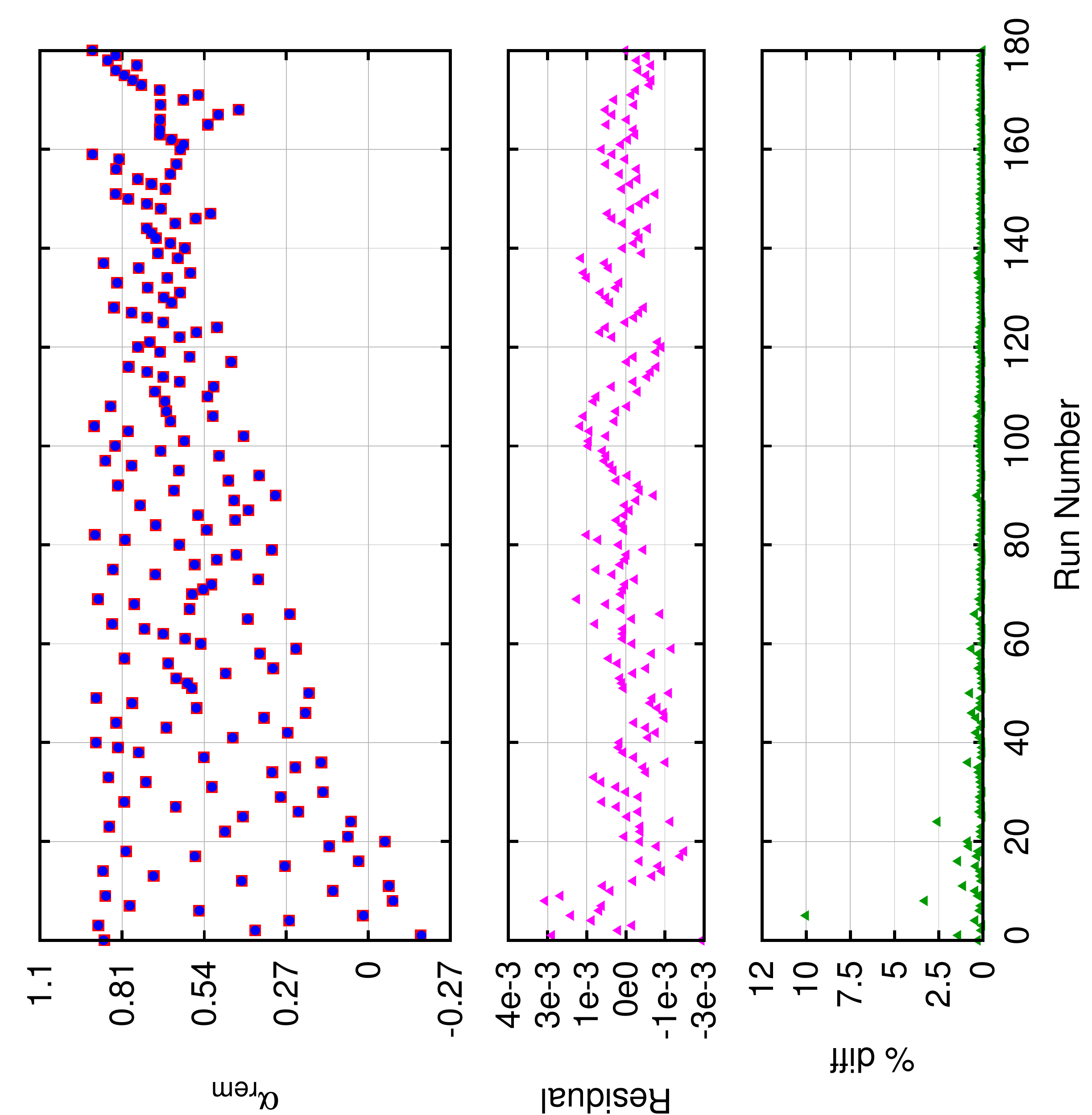}
\caption{Top panel: Red squares representing data and blue dots their
  corresponding fit. Bottom panels:
  Fitting residuals of the final remnant spin formula as given in
Eq.~(\ref{eq:4spin})
  (See also \cite{Healy:2016lce}).
\label{fig:spin}}
\end{figure}

In Fig.~\ref{fig:spin} we display the agreement between the
latest remnant spin formula as in Eq.~(\ref{eq:4spin})
(See also Ref.~\cite{Healy:2016lce}) with the updated set
of 181 simulations provided in this paper.
Table \ref{tab:fitparsms} gives the 19 parameters for the final spin fit.
Making use of the accurate determination of the final mass
via the isolated horizon formalism~\cite{Dreyer02a} we observe
that those coefficients with nearly vanishing values can be
adopted as precisely zero. Thus, Table \ref{tab:fitparsms_reduced} 
gives an alternative reduced set of 10 parameters fit.
As in the case of the final remnant mass, this reduced spin formula
may prove useful to extend these formulae to the precessing binaries case.
For instance, in Ref.~
\footnote{N.K. Johnson-McDaniel et al., LIGO Document T1600168, 
https://dcc.ligo.org/LIGO-T1600168/public} 
accurate results are found for the final spin
by augmenting the nonprecessing formulae with in-plane spins and 
spin evolution.

We find that comparing the residuals for the mass and spins between
the new and the previous fitting formulae implies a modest improvement,
i.e. the RMS for the new mass fit is 2.62396e-04 compared to
2.90334e-04 for the fit of Ref. \cite{Healy:2016lce}. While for
the final spin fit we find that the current RMS is 7.90772e-04 versus
8.15907e-04 for the Ref. \cite{Healy:2016lce}.

\subsection{Final Recoil Modeling}\label{sec:Recoil}

We model the total recoil as in Ref.~\cite{Healy:2014yta}
\begin{equation}\label{eq:empirical}
\vec{V}_{\rm recoil}(q,\vec\alpha_i)=v_m\,\hat{e}_1+
v_\perp(\cos(\xi)\,\hat{e}_1+\sin(\xi)\,\hat{e}_2),
\end{equation}
$\hat{e}_1,\hat{e}_2$ are orthogonal unit vectors in the
orbital plane, and $\xi$ measures the angle between the ``unequal mass''
and ``spin'' contributions to the recoil velocity in the orbital plane, and
with,
\begin{eqnarray}\label{eq:4recoil}
v_\perp &=& H\eta^2\left(\Dpar+ H_{2a} \Spar\dmt    
                     + H_{2b} \Dpar \Spar
                     + H_{3a} \Dpar^2\dmt\right.\nonumber\\
                     &&\left.+ H_{3b} \Spar^2\dmt
                     + H_{3c} \Dpar\Spar^2
                     + H_{3d} \Dpar^3\right.\nonumber\\
                     &&\left.+ H_{3e} \Dpar \dmt^2
                     + H_{4a} \Spar\Dpar^2\dmt
                     + H_{4b} \Spar^3 \dmt\right.\nonumber\\
                     &&\left.+ H_{4c} \Spar \dmt^3
                     + H_{4d} \Dpar \Spar \dmt^2\right.\nonumber\\
                     &&\left.+ H_{4e} \Dpar \Spar^3
                     + H_{4f} \Spar \Dpar^3\right),\nonumber\\
%\end{eqnarray}
%\begin{equation}\label{eq:xi}
\xi&=&a+b\, \Spar +c\, \dmt\Dpar.
%\end{equation}
\end{eqnarray}
Where 
\begin{equation}\label{eq:vm}
v_m=\eta^2 \delta m\left(A+B\,\delta m^2+C\,\delta{m}^4\right).
\end{equation}
and according to Ref.~\cite{Healy:2017mvh} we have $A=-8712\,km/s$,
and $B=-6516\,km/s$ and $C=3907\,km/s$.

\begin{figure}
\includegraphics[angle=270,width=0.99\columnwidth]{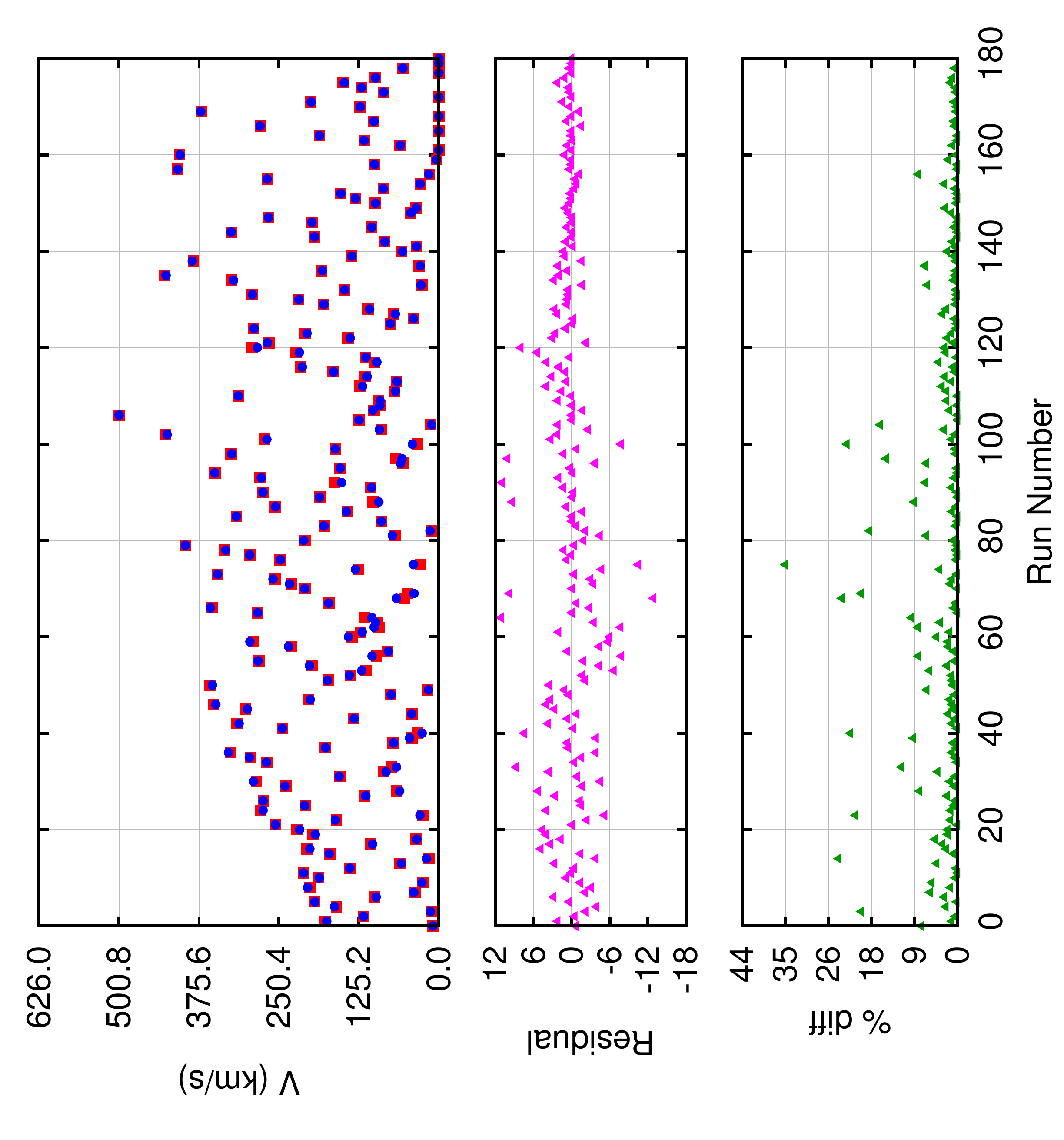}
\caption{Top panel: Red squares representing data and blue dots their
  corresponding fit to the in-plane recoil. Bottom panels:
  Fitting residuals of the final remnant recoil formula as given in
  Eq.~(\ref{eq:4recoil}) (See also \cite{Healy:2016lce}).
\label{fig:kick}}
\end{figure}

In Fig.~\ref{fig:kick} we display the agreement between the
recoil formula as in Eq.~(\ref{eq:4recoil})
(See also Ref.~\cite{Healy:2016lce}) with the updated set
of simulations provided in this paper.
Table \ref{tab:fitparsVL} provides the 17 parameters for the aligned
recoil formula.

In summary we find the fitting statistics as given in Table~\ref{tab:HL17stats}

\begin{table}
\caption{Fitting statistics for remnant formulae presented here.
}
\label{tab:HL17stats}
\begin{ruledtabular}
\begin{tabular}{cccc}
  Value  &	$M_{rem}/m$ &	$\alpha_{rem}$ & Recoil (km/s)\\
\hline
RMS&            2.62396e-04&    7.90772e-04&    3.48\\
Std. Dev.&      2.52011e-04&    7.58235e-04&    3.31\\
Avg. Diff.&	-6.38437e-06&	4.33099e-04&	0.21\\
Max Diff.&	1.19201e-03&	2.59799e-03&	10.98\\
Min Diff.&	-1.13027e-03&	-2.45274e-03&	-12.73
\end{tabular}
\end{ruledtabular}
\end{table}

\section{Further insight into the radiative and remnant relations}\label{sec:Correlations}

\begin{figure}
\includegraphics[angle=270,width=0.75\columnwidth]{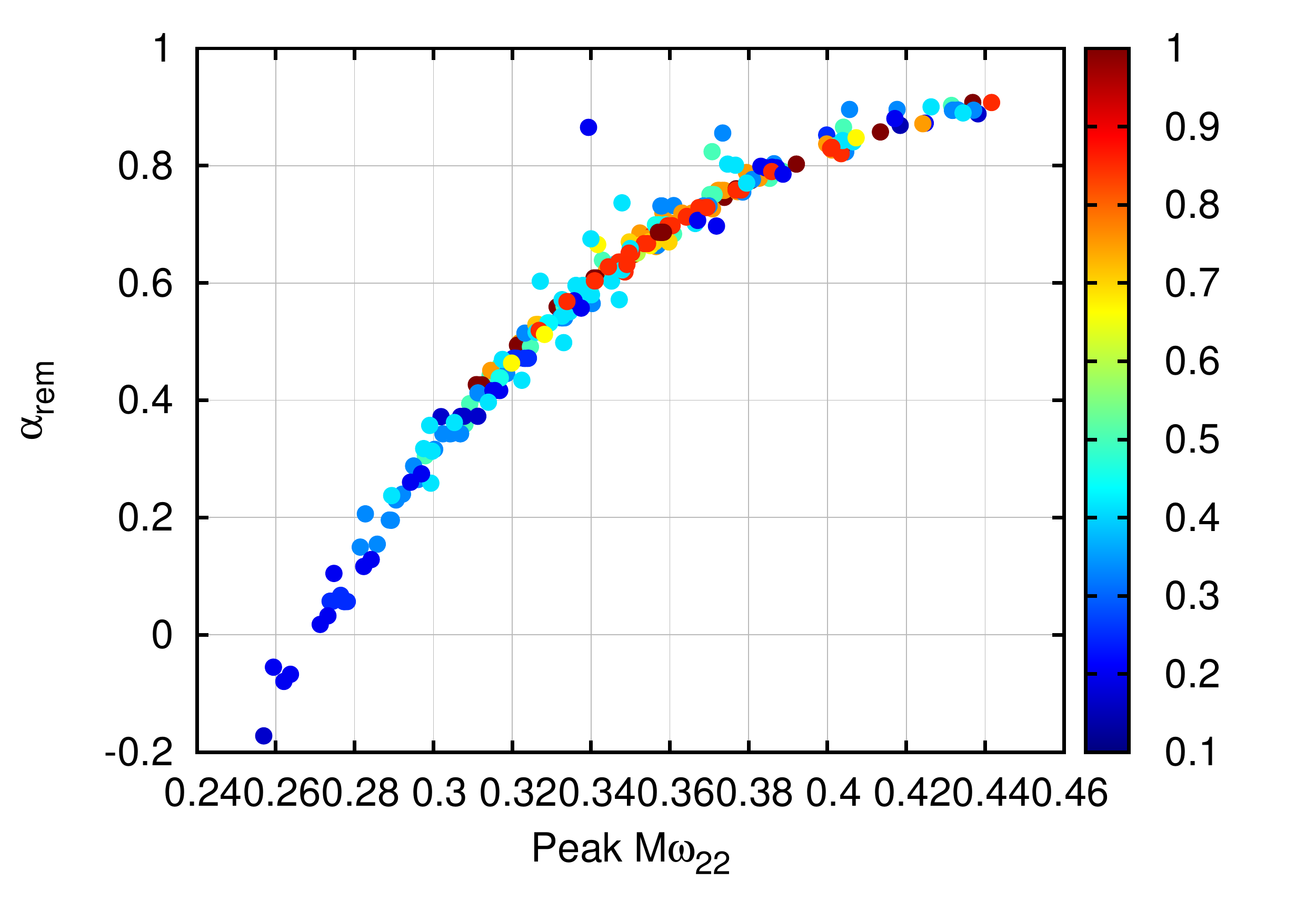}
\caption{The correlation between the final spin of the remnant and
  the (2,2)-mode frequency at the peak of the waveform amplitude over a range of spins and mass-ratios (in color). 
\label{fig:SOmegap}}
\end{figure}

We observe an interesting set of correlations among the remnant
and radiative merger variables as in Fig.~\ref{fig:SOmegap}.
Note that a similar correlation was pointed out independently
in Ref.~\cite{Healy:2014eua} between peak frequency and quasinormal modes of the
remnant black hole.

\begin{figure}
  \includegraphics[angle=270,width=0.75\columnwidth]{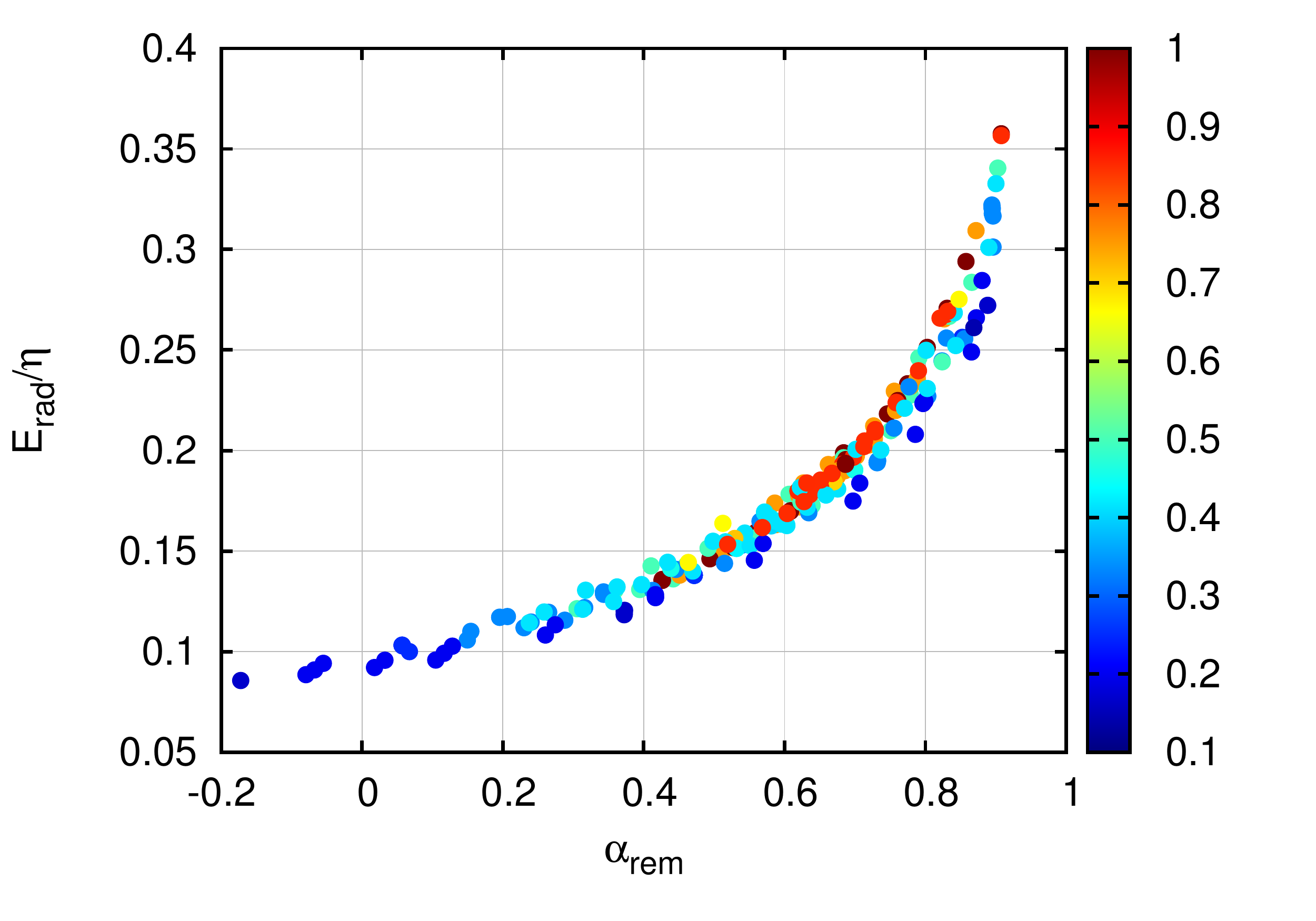}\\
  \includegraphics[angle=270,width=0.75\columnwidth]{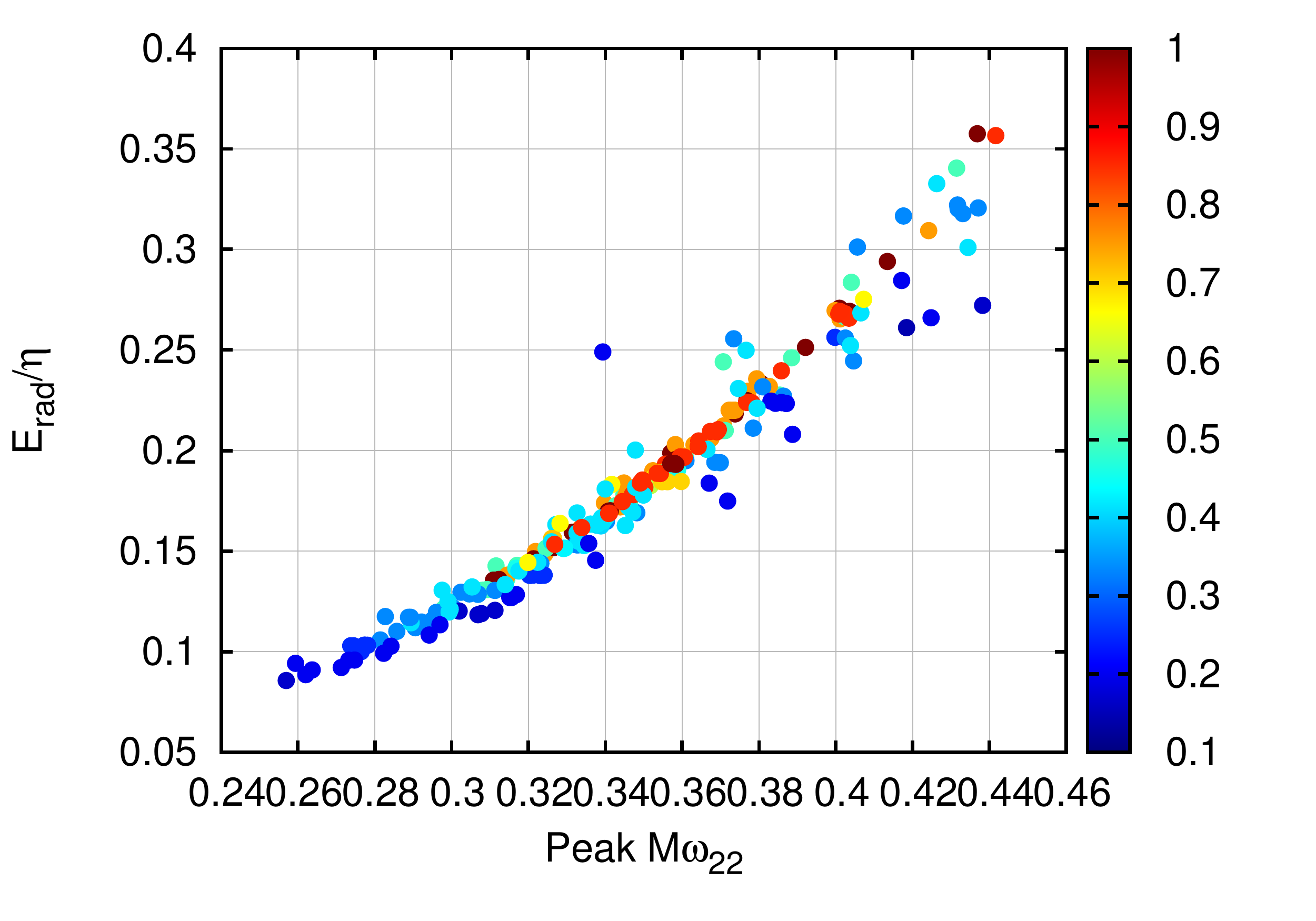}
\caption{The correlation between the normalized radiated energy and
  final spin of the remnant and
  the (2,2)-mode frequency at the peak of the waveform amplitude over a range of spins and mass-ratios (in color). 
\label{fig:ESO}}
\end{figure}

We also found interesting correlations for the normalized energy radiated,
$E_{rad}/\eta$, 
and the final spin or peak frequency displayed in Fig.~\ref{fig:ESO}.
Note the slight broadening of the correlation between Energy and peak
frequency for large values seems to be driven by three small mass ratio
simulations. This suggest this region of parameter space
should be supplemented with higher resolution and longer term simulations.

We note that we searched for other correlations among the peak, radiative,
and final remnant values and did not find simple and accurate relations as
those presented here for the three cases relating energy radiated, peak
frequency and final spin in an universal way, independent of the 
(moderate) mass ratios studied here.

We also note that these relationships, valid for our simulations of binary
black holes as governed by general relativity, could provide a test of
the theory of gravity when combined with independent observations of
such quantities by gravitational waves observatories.
For instance, excess power plots measuring wave amplitude in time vs. frequency
is commonly reported by LIGO analysis \cite{Chatterji:2004qg}.
Another example would be combining the measurement of the final mass 
and spin from a quasinormal modes (see for instance \cite{Baibhav:2017jhs})
with a measurement of the total mass and mass ratio from the inspiral waveform
to determine the total energy radiated.

\section{Conclusions and Discussion}\label{sec:Discussion}

%Hangup verified and modeled for q!=1
%Provided updated and new fits for formulae to apply to LIGO
%\begin{eqnarray}
%\frac12S_{0}&=&\left(\vec{S}\cdot\hat{L}+\frac{1}{2}\,\delta{m}\vec{\Delta}\cdot\hat{L}\right),\\
%\frac47S_{eff}&=&\left(\vec{S}\cdot\hat{L}+\frac{3}{7}\,\delta{m}\vec{\Delta}\cdot\hat{L}\right),\\
%\frac{188}{113}S_{PN}&=&\left(\vec{S}\cdot\hat{L}+\frac{75}{188}\,\delta{m}\vec{\Delta}\cdot\hat{L}\right),\\
%\frac32S_{hu}&=&\left(\vec{S}\cdot\hat{L}+\frac{1}{3}\,\delta{m}\vec{\Delta}\cdot\hat{L}\right).
%\end{eqnarray}
%Normalization such that in the particle limit we obtain $\vec{S}_2$.
%Alternatively

We have performed 74 new simulations of unequal mass, spinning nonprecessing 
binary black holes to investigate the dynamics of their late inspiral and
merger leading find that the hangup phenomena (acceleration/deceleration of
merger with respect to the nonspinning case) can be represented at leading
spin order by the following quantity
\beq
S_{hu}=\left((1+\frac{1}{2q})\,\vec{S}_1+(1+\frac{1}{2}q)\,\vec{S}_2\right)\cdot\hat{L}.
\eeq

We observed a clear better match with respect to alternative effective 
description used in the modeling and post-Newtonian descriptions
\begin{eqnarray}
S_{0}&=&\left((1+\frac1q)\,\vec{S}_1+(1+q)\,\vec{S}_2\right)\cdot\hat{L},\\
S_{eff}&=&\left((1+\frac{3}{4q})\,\vec{S}_1+(1+\frac{3}{4}q)\,\vec{S}_2\right)\cdot\hat{L},\\
S_{PN}&=&\left((1+\frac{75}{113q})\,\vec{S}_1+(1+\frac{75}{113}q)\,\vec{S}_2\right)\cdot\hat{L}.
\end{eqnarray}

\begin{figure}
  \includegraphics[width=0.75\columnwidth]{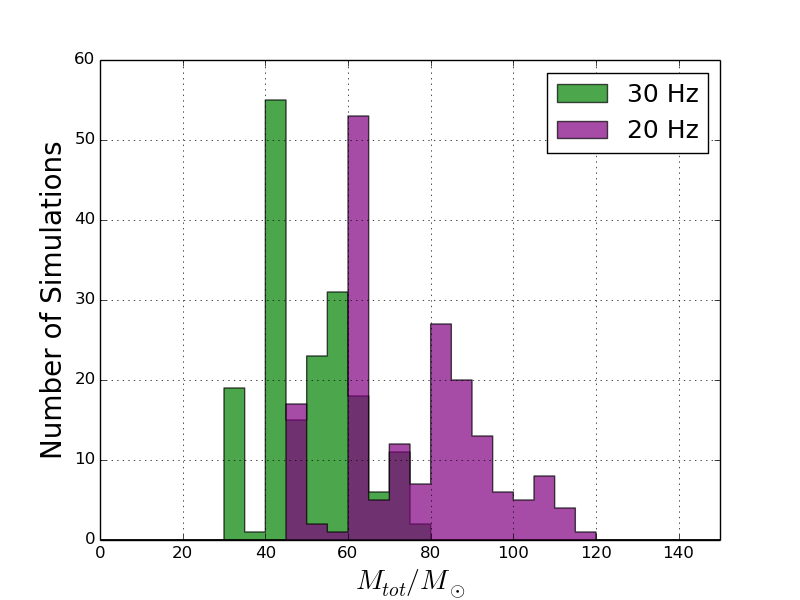}
  \includegraphics[width=0.75\columnwidth]{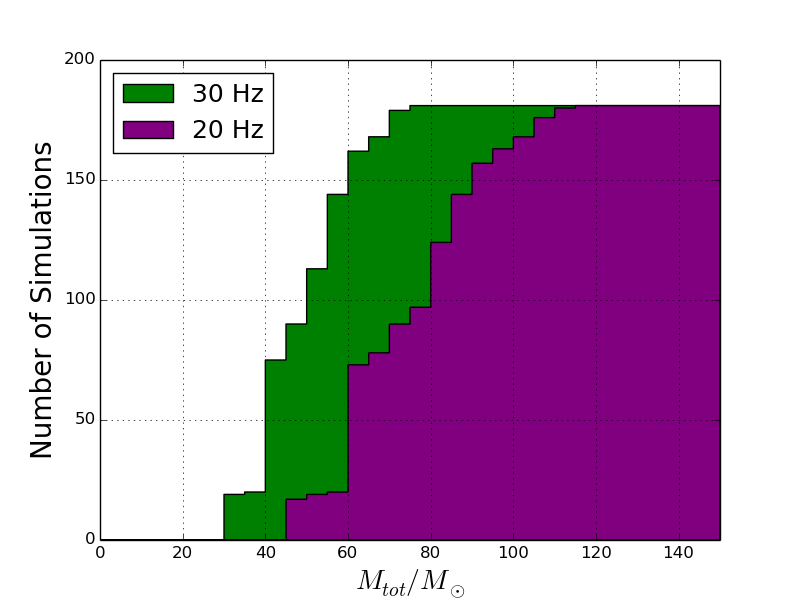}
  \caption{The lowest mass representation of the simulated waveforms for a
    starting frequency of 20Hz and 30Hz at the source frame. On top, the number
    of waveforms in individual bins of $5M_\odot$ and on the bottom, the cumulative number of waveforms. 
\label{fig:mtotal}}
\end{figure}

We have also generated new simulations to add to the waveform data bank available for parameter estimation of gravitational wave observations
\cite{Abbott:2016apu,Lange:2017wki},
and soon to be included in a new release of the RIT waveform Catalog
\cite{Healy:2017psd}
(\url{http://ccrg.rit.edu/~RITCatalog/}). For instance,
the lowest mass coverage of our full set 
of aligned runs used here is summarized in Fig.~\ref{fig:mtotal}

\begin{figure}
  \includegraphics[angle=270,width=0.75\columnwidth]{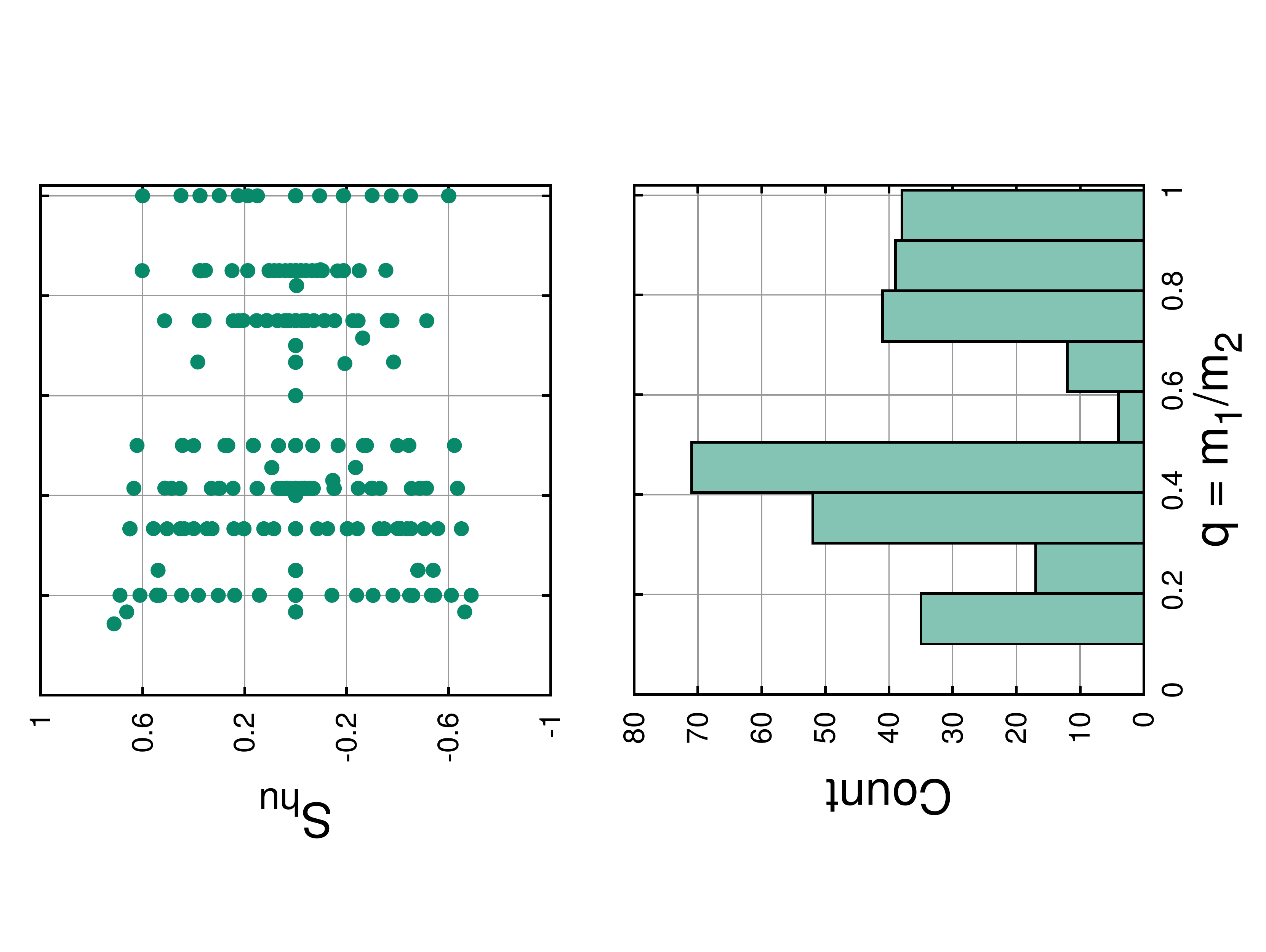}
  \caption{Coverage of the whole set of aligned spinning binary black 
    simulations used in this paper.
\label{fig:chieff}}
\end{figure}

Our simulations and results can also be used to 
improve ``phenomenological models'' 
by considering the use of $S_{hu}$ as the variable for aligned binaries
instead of $\chi_{eff}=\tilde{S}_0$ or $\chi_{PN}=\tilde{S}_{PN}$
(as used in \cite{Khan:2015jqa}). 
We display the coverage in this variable of our whole
set of simulations from this paper and Refs.\cite{Healy:2014eua} and \cite{Healy:2016lce}
in Fig.~\ref{fig:chieff}. We observe that there is a lack of simulations
in the regions of high spins (aligned or counteraligned) and the region
of small mass ratios. While there are some simulations available in those
regions we have not included them in this particular set of studies until
we have a systematic coverage of those regions.

Finally, the new fits to the whole set of simulations, particularly 
improve the accuracy of the remnant recoil and peak luminosity, 
amplitude and frequency for applications to the observations of
gravitational waves and tests of general relativity.

%%%%%%%%%%%%%%%%%%%%%%%%%%%%%%%%%%%%%%%%%%%%%
\acknowledgments 
The authors thank M. Campanelli, N.K.Johnson-McDaniel, D.Keitel,
H. Nakano, R. O'Shaughnessy and Y. Zlochower for discussions on this work.
The authors gratefully acknowledge the National Science Foundation (NSF)
for financial support from Grants
No.\ PHY-1607520, No.\ PHY-1707946, No.\ ACI-1550436, No.\ AST-1516150,
No.\ ACI-1516125, No.\ PHY-1726215.
This work used the Extreme Science and Engineering
Discovery Environment (XSEDE) [allocation TG-PHY060027N], which is
supported by NSF grant No. ACI-1548562.
Computational resources were also provided by the NewHorizons and
BlueSky Clusters at the Rochester Institute of Technology, which were
supported by NSF grants No.\ PHY-0722703, No.\ DMS-0820923, No.\
AST-1028087, and No.\ PHY-1229173. 
%%%%%%%%%%%%%%%%%%%%%%%%%%%%%%%%%%%%%%%

%\newpage
\clearpage
\appendix

\section{Tables of initial data and results of the new simulations}\label{app:ID}

In this appendix we provide the tables of the initial data 
(Table \ref{tab:ID}) used to start
the full numerical evolutions and a Table \ref{tab:IDr} with
the mass and spin parameters after 
they settle into a more physical value from the initial conformal
flatness mathematical choice by radiating it away 
(a fiducial $t=200M$, within an orbit from start).

We also provide a Table \ref{tab:ecc} with the initial orbital frequency
and eccentricity as well as the number of orbits to merger and the
final eccentricity, expected to be reduced from its initial value by
gravitational radiation, at a rate proportional to
$d^{19/12}$ according to \cite{Peters:1964zz}, with $d$, the separation of the
binary (see, for instance, Fig.~6 of Ref.~\cite{Mroue:2010re} or
Fig.~9 in Ref.~\cite{Lousto:2015uwa}).

We provide a Table \ref{tab:spinerad}
with the values of the energy radiated during the simulation
and the final black hole spin as measured through the (most accurate)
isolated horizon formalism \cite{Dreyer02a}.
A Table \ref{tab:kicks}
with the recoil velocity completes the properties of the remnant black hole.

Finally, Tables \ref{tab:peakOA}, provide the data of
the peak amplitude and frequency of the gravitational wave strain of
the (2,2) modes for the whole set of 181 simulations.

%\clearpage
\begin{longtable*}{lccccccccccccc}
\caption{Initial data parameters for the quasi-circular
configurations with a smaller mass black hole (labeled 1),
and a larger mass spinning black hole (labeled 2). The punctures are located
at $\vec r_1 = (x_1,0,0)$ and $\vec r_2 = (x_2,0,0)$, with momenta
$P=\pm (P_r, P_t,0)$, spins $\vec S_i = (0, 0, S_i)$, mass parameters
$m^p/m$, horizon (Christodoulou) masses $m^H/m$, total ADM mass
$M_{\rm ADM}$, and dimensionless spins $a/m_H = S/m_H^2$.
The configuration are denoted by QX\_Y\_Z, where X gives the mass
ratio $m^H_1 / m^H_2$, Y gives the  spin of the smaller BH
($a_1/m_H^2$), and Z
gives the spin of the larger BH $(a_2/m_H^2)$.}
\label{tab:ID}\\
\hline
\hline
Run   & $x_1/m$ & $x_2/m$  & $P_r/m$    & $P_t/m$ & $m^p_1/m$ & $m^p_2/m$ & $S_1/m^2$ & $S_2/m^2$ & $m^H_1/m$ & $m^H_2/m$ & $M_{\rm ADM}/m$ & $a_1/m_1^H$ & $a_2/m_2^H$\\
\hline
\endfirsthead

\multicolumn{14}{c}
{{\tablename\ \thetable{} -- continued from previous
page}}\\
\hline
Run   & $x_1/m$ & $x_2/m$  & $P_r/m$    & $P_t/m$ & $m^p_1/m$ & $m^p_2/m$ & $S_1/m^2$ & $S_2/m^2$ & $m^H_1/m$ & $m^H_2/m$ & $M_{\rm ADM}/m$ & $a_1/m_1^H$ & $a_2/m_2^H$\\
\hline
\endhead
\hline \multicolumn{14}{r}{{Continued on next page}} \\
\hline
\endfoot
\hline\hline
\endlastfoot

1 & -7.88 & 1.12 & -2.14e-04 & 0.04251 & 0.07322 & 0.5405 & 0.0125 & 0.6125 & 0.125 & 0.875 & 0.995 & 0.8 & 0.8 \\
2 & -10.83 & 2.17 & -1.47e-04 & 0.04609 & 0.1602 & 0.7352 & 0 & -0.3472 & 0.1667 & 0.8333 & 0.9955 & 0 & -0.5 \\
3 & -10.83 & 2.17 & -1.17e-04 & 0.04393 & 0.1603 & 0.7352 & 0 & 0.3472 & 0.1667 & 0.8333 & 0.9953 & 0 & 0.5 \\
4 & -10.83 & 2.17 & -1.40e-04 & 0.0457 & 0.1423 & 0.8065 & -0.01389 & -0.1736 & 0.1667 & 0.8333 & 0.9955 & -0.5 & -0.25 \\
5 & -10.83 & 2.17 & -1.25e-04 & 0.0446 & 0.1423 & 0.8065 & -0.01389 & 0.1736 & 0.1667 & 0.8333 & 0.9954 & -0.5 & 0.25 \\
6 & -10.83 & 2.17 & -1.36e-04 & 0.04534 & 0.1423 & 0.8065 & 0.01389 & -0.1736 & 0.1667 & 0.8333 & 0.9955 & 0.5 & -0.25 \\
7 & -10.83 & 2.17 & -1.21e-04 & 0.04428 & 0.1423 & 0.8065 & 0.01389 & 0.1736 & 0.1667 & 0.8333 & 0.9954 & 0.5 & 0.25 \\
8 & -10.83 & 2.17 & -1.55e-04 & 0.04662 & 0.1422 & 0.655 & -0.01389 & -0.4514 & 0.1667 & 0.8333 & 0.9956 & -0.5 & -0.65 \\
\end{longtable*}

\clearpage
\begin{longtable*}{llcccccccc}
\caption{The mass and spin of the BHBs in Table~\ref{tab:ID} after the
BHs had time to equilibrate ($t/m=200$).}
\label{tab:IDr}\\
\hline
\hline
Run & Config.  & $q^r$ & $m^r_1/m$ & $m^r_2/m$ & $\alpha^r_1$ & $\alpha^r_2$ & $\delta m_r$ & $S_r/m^2_r$ & $\Delta_r/m^2_r$ \\
\hline
\endfirsthead

\multicolumn{10}{c}
{{\tablename\ \thetable{} -- continued from previous page}}\\
\hline
Run & Config.  & $q^r$ & $m^r_1/m$ & $m^r_2/m$ & $\alpha^r_1$ & $\alpha^r_2$ & $\delta m_r$ & $S_r/m^2_r$ & $\Delta_r/m^2_r$ \\
\hline
\endhead
\hline \multicolumn{10}{r}{{Continued on next page}} \\
\hline
\endfoot
\hline\hline
\endlastfoot

1 & Q0.1429\_0.8000\_0.8000 &	$0.142939$ & $0.125001$ & $0.874504$ & $0.800504$ & $0.800939$ & $-0.749875$ & $0.625652$ & $0.600659$\\
2 & Q0.2000\_0.0000\_-0.5000 &	$0.200013$ & $0.166669$ & $0.833287$ & $-0.000002$ & $-0.500061$ & $-0.666648$ & $-0.347257$ & $-0.416713$\\
3 & Q0.2000\_0.0000\_0.5000 &	$0.200014$ & $0.166669$ & $0.833287$ & $-0.000000$ & $0.500060$ & $-0.666647$ & $0.347256$ & $0.416712$\\
4 & Q0.2000\_-0.5000\_-0.2500 &	$0.200012$ & $0.166676$ & $0.833332$ & $-0.500066$ & $-0.250017$ & $-0.666651$ & $-0.187512$ & $-0.124997$\\
5 & Q0.2000\_-0.5000\_0.2500 &	$0.200011$ & $0.166676$ & $0.833332$ & $-0.500081$ & $0.250019$ & $-0.666651$ & $0.159728$ & $0.291698$\\
6 & Q0.2000\_0.5000\_-0.2500 &	$0.200012$ & $0.166676$ & $0.833331$ & $0.500052$ & $-0.250019$ & $-0.666650$ & $-0.159730$ & $-0.291693$\\
7 & Q0.2000\_0.5000\_0.2500 &	$0.200011$ & $0.166675$ & $0.833330$ & $0.500055$ & $0.250020$ & $-0.666651$ & $0.187513$ & $0.125002$\\
8 & Q0.2000\_-0.5000\_-0.6500 &	$0.200049$ & $0.166677$ & $0.833178$ & $-0.500052$ & $-0.650269$ & $-0.666599$ & $-0.465435$ & $-0.458509$\\
\end{longtable*}

\clearpage
\begin{longtable*}{llcccc}
\caption{Table of the initial orbital frequency $m\omega_i$,
number of orbits to merger, $N$, and the initial and final eccentricities,
$e_i$ and $e_f$ for the spinning cases.}
\label{tab:ecc}\\
\hline
\hline
Run & Config.   & $m\omega_i$ & $N$ & $e_i$ & $e_f$ \\
\hline
\endfirsthead

\multicolumn{6}{c}
{{\tablename\ \thetable{} -- continued from previous page}}\\
\hline
Run & Config.   & $m\omega_i$ & $N$ & $e_i$ & $e_f$ \\
\hline
\endhead
\hline \multicolumn{6}{r}{{Continued on next page}} \\
\hline
\endfoot
\hline\hline
\endlastfoot

1 & Q0.1429\_0.8000\_0.8000 &	$0.0311$ & $16.3$ & $0.0020$ & $0.0010$\\
2 & Q0.2000\_0.0000\_-0.5000 &	$0.0193$ & $15.2$ & $0.0046$ & $0.0007$\\
3 & Q0.2000\_0.0000\_0.5000 &	$0.0188$ & $24.7$ & $0.0025$ & $0.0003$\\
4 & Q0.2000\_-0.5000\_-0.2500 &	$0.0192$ & $16.6$ & $0.0049$ & $0.0006$\\
5 & Q0.2000\_-0.5000\_0.2500 &	$0.0189$ & $21.4$ & $0.0042$ & $0.0005$\\
6 & Q0.2000\_0.5000\_-0.2500 &	$0.0191$ & $17.8$ & $0.0023$ & $0.0006$\\
7 & Q0.2000\_0.5000\_0.2500 &	$0.0189$ & $22.7$ & $0.0024$ & $0.0006$\\
8 & Q0.2000\_-0.5000\_-0.6500 &	$0.0194$ & $13.2$ & $0.0029$ & $0.0009$\\
\end{longtable*}

\clearpage
\begin{longtable*}{llcc}
\caption{The final energy radiated and spin as measured using the IH formalism
The error bars are due to variations in the measured mass and spin with time.}
\label{tab:spinerad}\\
\hline
\hline
Run & Config. & $\delta \mathcal{M}^{IH}$ & $\alpha_{\mathrm{rem}}^{IH}$ \\
\hline
\hline
\endfirsthead

\multicolumn{4}{c}
{{\tablename\ \thetable{} -- continued from previous page}}\\
\hline
Run & Config. & $\delta \mathcal{M}^{IH}$ & $\alpha_{\mathrm{rem}}^{IH}$ \\
\hline
\endhead
\hline \multicolumn{4}{r}{{Continued on next page}} \\
\hline
\endfoot
\hline\hline
\endlastfoot

1 & Q0.1429\_0.8000\_0.8000 &	 $0.028570 \pm 0.000022$ & 	$0.868808 \pm 0.001039$ \\
2 & Q0.2000\_0.0000\_-0.5000 &	 $0.013777 \pm 0.000000$ & 	$0.116641 \pm 0.000009$ \\
3 & Q0.2000\_0.0000\_0.5000 &	 $0.025525 \pm 0.000002$ & 	$0.706681 \pm 0.000034$ \\
4 & Q0.2000\_-0.5000\_-0.2500 &	 $0.015035 \pm 0.000001$ & 	$0.260381 \pm 0.000005$ \\
5 & Q0.2000\_-0.5000\_0.2500 &	 $0.020200 \pm 0.000001$ & 	$0.557126 \pm 0.000003$ \\
6 & Q0.2000\_0.5000\_-0.2500 &	 $0.015740 \pm 0.000002$ & 	$0.274468 \pm 0.000013$ \\
7 & Q0.2000\_0.5000\_0.2500 &	 $0.021363 \pm 0.000002$ & 	$0.569568 \pm 0.000002$ \\
8 & Q0.2000\_-0.5000\_-0.6500 &	 $0.012792 \pm 0.000006$ & 	$0.017695 \pm 0.000001$ \\
\end{longtable*}

\clearpage
\begin{longtable*}{llrrcc}
\caption{The recoil velocity (in km/s) and peak luminosity as calculated using $\ell_{max}=6$ and
$r_{max}=113.0m$ for spinning systems.
$W_{V}$ and $W_{\mathcal{L}}$ are the weights used in the least-squares fitting.}
% The error estimates are detailed in Appendix~\ref{app:errors}.}
\label{tab:kicks}\\
\hline
\hline
Run & Config. &  $V$ [km/s] & $W_{V}$ [km/s] & $\mathcal{L}_{\text{max}}$ & $W_{\mathcal{L}}$\\
\hline
\endfirsthead

\multicolumn{5}{c}
{{\tablename\ \thetable{} -- continued from previous page}}\\
\hline
Run & Config. &  $V$ [km/s] & $W_{V}$ [km/s] & $\mathcal{L}_{\text{max}}$ & $W_{\mathcal{L}}$\\
\hline
\endhead
\hline \multicolumn{6}{r}{{Continued on next page}} \\
\hline
\endfoot
\hline\hline
\endlastfoot

1 & Q0.1429\_0.8000\_0.8000 &	$9.00 \pm 6.89$ & 6.77 & 3.4148e-04 $\pm$ 3.9140e-05 & 1.0600e-05 \\
2 & Q0.2000\_0.0000\_-0.5000 &	$188.07 \pm 6.77$ & 5.53 & 2.2478e-04 $\pm$ 2.3403e-05 & 2.3097e-05 \\
3 & Q0.2000\_0.0000\_0.5000 &	$61.54 \pm 11.35$ & 5.53 & 4.0031e-04 $\pm$ 2.8094e-05 & 2.3097e-05 \\
4 & Q0.2000\_-0.5000\_-0.2500 &	$159.44 \pm 8.51$ & 5.53 & 2.4906e-04 $\pm$ 2.3609e-05 & 2.3097e-05 \\
5 & Q0.2000\_-0.5000\_0.2500 &	$101.21 \pm 11.69$ & 5.53 & 3.3109e-04 $\pm$ 2.5973e-05 & 2.3097e-05 \\
6 & Q0.2000\_0.5000\_-0.2500 &	$170.47 \pm 8.64$ & 5.53 & 2.5227e-04 $\pm$ 2.3526e-05 & 2.3097e-05 \\
7 & Q0.2000\_0.5000\_0.2500 &	$107.36 \pm 11.25$ & 5.53 & 3.3546e-04 $\pm$ 2.5964e-05 & 2.3097e-05 \\
8 & Q0.2000\_-0.5000\_-0.6500 &	$194.57 \pm 6.39$ & 5.53 & 2.1025e-04 $\pm$ 2.3236e-05 & 2.3097e-05 \\
\end{longtable*}

\clearpage
\begin{longtable*}{llrrcc}
\caption{The peak gravitatonal wave frequency and amplitude of the 22 strain mode for the most recent 
runs presented in this paper.  }
\label{tab:peakOA}\\
\hline
\hline
Run & Config. &  $m\omega_{22}^{peak}$ &  $(r/m)|h_{22}^{peak}|$ \\
\hline
\hline
\endfirsthead

\multicolumn{5}{c}
{{\tablename\ \thetable{} -- continued from previous page}}\\
\hline
Run & Config. &  $m\omega_{22}^{peak}$ &  $(r/m)|h_{22}^{peak}|$ \\
\hline
\endhead
\hline \multicolumn{6}{r}{{Continued on next page}} \\
\hline
\endfoot
\hline\hline
\endlastfoot

1 & Q0.1429\_0.8000\_0.8000 &	$0.4184 \pm 0.0038$ & 0.1650 $\pm$ 0.0005 \\
2 & Q0.2000\_0.0000\_-0.5000 &	$0.2823 \pm 0.0029$ & 0.2049 $\pm$ 0.0002 \\
3 & Q0.2000\_0.0000\_0.5000 &	$0.3670 \pm 0.0052$ & 0.2110 $\pm$ 0.0008 \\
4 & Q0.2000\_-0.5000\_-0.2500 &	$0.2942 \pm 0.0025$ & 0.2065 $\pm$ 0.0010 \\
5 & Q0.2000\_-0.5000\_0.2500 &	$0.3375 \pm 0.0025$ & 0.2088 $\pm$ 0.0014 \\
6 & Q0.2000\_0.5000\_-0.2500 &	$0.2970 \pm 0.0044$ & 0.2060 $\pm$ 0.0009 \\
7 & Q0.2000\_0.5000\_0.2500 &	$0.3357 \pm 0.0030$ & 0.2094 $\pm$ 0.0016 \\
8 & Q0.2000\_-0.5000\_-0.6500 &	$0.2712 \pm 0.0030$ & 0.2038 $\pm$ 0.0001 \\
\end{longtable*}

\clearpage
\begin{longtable*}{llrrcc}
\caption{The peak gravitatonal wave frequency and amplitude of the 22 strain mode for the  
runs presented in Ref~\cite{Healy:2016lce}  }
\label{tab:peakOA2}\\
\hline
\hline
Run & Config. &  $m\omega_{22}^{peak}$ & $(r/m)|h_{22}^{peak}|$ \\
\hline
\endfirsthead

\multicolumn{6}{c}
{{\tablename\ \thetable{} -- continued from previous page}}\\
\hline
Run & Config. &  $m\omega_{22}^{peak}$ & $(r/m)|h_{22}^{peak}|$ \\
\hline
\endhead
\hline \multicolumn{6}{r}{{Continued on next page}} \\
\hline
\endfoot
\hline\hline
\endlastfoot

75 & Q0.1667\_0.0000\_0.0000 &	$0.3078 \pm 0.0025$ & 0.1812 $\pm$ 0.0004 \\
76 & Q0.2000\_0.0000\_0.0000 &	$0.3152 \pm 0.0028$ & 0.2074 $\pm$ 0.0005 \\
77 & Q0.2500\_0.0000\_0.0000 &	$0.3205 \pm 0.0033$ & 0.2406 $\pm$ 0.0005 \\
78 & Q0.3333\_0.0000\_0.0000 &	$0.3328 \pm 0.0034$ & 0.2859 $\pm$ 0.0002 \\
79 & Q0.3333\_0.0000\_-0.5000 &	$0.3003 \pm 0.0036$ & 0.2836 $\pm$ 0.0000 \\
80 & Q0.3333\_0.0000\_0.5000 &	$0.3785 \pm 0.0024$ & 0.2878 $\pm$ 0.0001 \\
81 & Q0.3333\_0.0000\_-0.8500 &	$0.2857 \pm 0.0045$ & 0.2806 $\pm$ 0.0002 \\
82 & Q0.3333\_0.0000\_0.8500 &	$0.4056 \pm 0.0072$ & 0.2891 $\pm$ 0.0003 \\
\end{longtable*}

\begin{table*}
\caption{The peak gravitatonal wave frequency and amplitude of the 22 strain mode for the 
runs presented in Ref~\cite{Healy:2014yta}.  }
\label{tab:peakOA1}
\begin{ruledtabular}
\begin{tabular}{llrrcc}
Run & Config. &  $m\omega_{22}^{peak}$ & $(r/m)|h_{22}^{peak}|$ \\
\hline

146 & Q1.00\_0.00\_0.00 &	$0.3580 \pm 0.0080$ & 0.3937 $\pm$ 0.0029 \\
147 & Q1.00\_0.00\_0.40 &	$0.3738 \pm 0.0031$ & 0.3935 $\pm$ 0.0002 \\
148 & Q1.00\_0.00\_0.60 &	$0.3806 \pm 0.0043$ & 0.3935 $\pm$ 0.0005 \\
149 & Q1.00\_0.00\_0.80 &	$0.3921 \pm 0.0031$ & 0.3937 $\pm$ 0.0005 \\
150 & Q1.00\_0.20\_0.80 &	$0.4010 \pm 0.0046$ & 0.3948 $\pm$ 0.0009 \\
151 & Q1.00\_-0.40\_0.40 &	$0.3584 \pm 0.0032$ & 0.3934 $\pm$ 0.0002 \\
152 & Q1.00\_0.40\_0.80 &	$0.4134 \pm 0.0032$ & 0.3951 $\pm$ 0.0010 \\
\hline
\end{tabular}
\end{ruledtabular}
\end{table*}

\section{Tables of fitting parameters}\label{app:fit}

Here we provide the values for the 19 (or 17) fitting parameters needed to
represent the fourth order expansion of the remnant and radiation quantities
we model.
Table \ref{tab:fitparsms}
give the 19 parameters for the final mass, Eq.~(\ref{eq:4mass})
and spin, Eq.~(\ref{eq:4spin})
formulae. Table \ref{tab:fitparsms_reduced} 
gives a reduced set of 9 and 10 parameters
fit making use of the accurate determination of the final mass and spin
via the isolated horizon formalism~\cite{Dreyer02a}.  The residuals for
these reduced fits, while not as low as the full fit, are comparable.  
The mass fit RMS increases to 4.4e-4 from 2.6e-4 and the spin fit RMS
increases to 9.3e-4 from 7.9e-4.
This may provide helpful in a hierarchical approach to extend these
formulae to precessing binaries.

Table \ref{tab:fitparsVL} provides the 17 parameters for the aligned
recoil formula, Eq.~(\ref{eq:4recoil})
and the 19 of the peak luminosity, Eq.~(\ref{eq:4plum}). 
Table \ref{tab:fitparsOA} completes the fourth order parameterization
of the peak strain amplitude and frequency used in 
Eq.~(\ref{eq:PAmp}) and Eq.~(\ref{eq:Pfreq}).

\begin{table*}
\caption{
Table of fitting parameters for the mass, and spin formulas.}\label{tab:fitparsms}
\begin{ruledtabular}
\begin{tabular}{lr|lr}
M0  & $ 0.951714 \pm 0.000019$ &	L0  & $ 0.686786 \pm 0.000019$ \\
K1  & $-0.052203 \pm 0.000129$ &	L1  & $ 0.614468 \pm 0.000125$ \\
K2a & $-0.005305 \pm 0.000232$ &	L2a & $-0.149948 \pm 0.000249$ \\
K2b & $-0.061114 \pm 0.000416$ &	L2b & $-0.115787 \pm 0.000417$ \\
K2c & $-0.001567 \pm 0.000116$ &	L2c & $-0.004314 \pm 0.000108$ \\
K2d & $ 1.995914 \pm 0.000235$ &	L2d & $ 0.800085 \pm 0.000228$ \\
K3a & $-0.003966 \pm 0.001365$ &	L3a & $-0.073908 \pm 0.001334$ \\
K3b & $-0.005392 \pm 0.000618$ &	L3b & $-0.011940 \pm 0.000717$ \\
K3c & $-0.110043 \pm 0.000980$ &	L3c & $-0.079447 \pm 0.000956$ \\
K3d & $ 0.015735 \pm 0.000855$ &	L3d & $ 1.546260 \pm 0.000886$ \\
K4a & $-0.038715 \pm 0.002467$ &	L4a & $-0.038602 \pm 0.002548$ \\
K4b & $-0.001674 \pm 0.000547$ &	L4b & $-0.003690 \pm 0.000658$ \\
K4c & $-0.000351 \pm 0.000146$ &	L4c & $ 0.000511 \pm 0.000134$ \\
K4d & $-0.157569 \pm 0.002262$ &	L4d & $-0.056376 \pm 0.002168$ \\
K4e & $ 0.009310 \pm 0.001646$ &	L4e & $-0.001008 \pm 0.000340$ \\
K4f & $ 2.977562 \pm 0.000601$ &	L4f & $ 0.958901 \pm 0.000610$ \\
K4g & $ 0.001792 \pm 0.000712$ &	L4g & $-0.107740 \pm 0.001174$ \\
K4h & $-0.004809 \pm 0.000972$ &	L4h & $-0.016576 \pm 0.001058$ \\
K4i & $ 0.084504 \pm 0.001929$ &	L4i & $-0.082960 \pm 0.001991$ \\
\end{tabular}
\end{ruledtabular}
\end{table*}

\begin{table*}
\caption{
Table of fitting parameters for the mass and spin formulas using a reduced
number of fitting parameters.}\label{tab:fitparsms_reduced}
\begin{ruledtabular}
\begin{tabular}{lr|lr}
M0  & $ 0.951432 \pm 0.000014$ &        L0  & $ 0.685913 \pm 0.000014$ \\
K1  & $-0.052209 \pm 0.000077$ &        L1  & $ 0.613022 \pm 0.000092$ \\
K2a & $                     0$ &        L2a & $-0.148075 \pm 0.000174$ \\
K2b & $-0.060308 \pm 0.000349$ &        L2b & $-0.102671 \pm 0.000348$ \\
K2c & $                     0$ &        L2c & $                     0$ \\
K2d & $ 1.996335 \pm 0.000210$ &        L2d & $ 0.806511 \pm 0.000206$ \\
K3a & $                     0$ &        L3a & $                     0$ \\
K3b & $                     0$ &        L3b & $                     0$ \\
K3c & $-0.108377 \pm 0.000612$ &        L3c & $-0.074281 \pm 0.000598$ \\
K3d & $ 0.038011 \pm 0.000376$ &        L3d & $ 1.556791 \pm 0.000684$ \\
K4a & $                     0$ &        L4a & $                     0$ \\
K4b & $                     0$ &        L4b & $                     0$ \\
K4c & $                     0$ &        L4c & $                     0$ \\
K4d & $-0.154938 \pm 0.001817$ &        L4d & $-0.086944 \pm 0.001545$ \\
K4e & $                     0$ &        L4e & $                     0$ \\
K4f & $ 2.977785 \pm 0.000568$ &        L4f & $ 0.948992 \pm 0.000553$ \\
K4g & $                     0$ &        L4g & $-0.110623 \pm 0.000940$ \\
K4h & $                     0$ &        L4h & $                     0$ \\
K4i & $ 0.082810 \pm 0.001171$ &        L4i & $                     0$ \\
\end{tabular}
\end{ruledtabular}
\end{table*}

\begin{table*}
\caption{
Table of fitting parameters (left) for the recoil (in Km/s)
and (right) peak luminosity formulas. Nonspinning coefficients
N0, N2d, and N4f were determined in Ref.~\cite{Healy:2017mvh}. }\label{tab:fitparsVL}
\begin{ruledtabular}
\begin{tabular}{lr|lr}
H   & $ 7499.115 \pm 9.244136$ &	N0  & $ 1.026e-03 \pm 1.727e-6$ \\
H2a & $-1.736510 \pm 0.032585$ &	N1  & $ 8.839321e-04 \pm 4.914069e-06$ \\
H2b & $-0.598144 \pm 0.014548$ &	N2a & $ 1.076865e-04 \pm 1.141520e-05$ \\
H3a & $-0.318117 \pm 0.032373$ &	N2b & $ 6.882092e-04 \pm 1.082919e-05$ \\
H3b & $-0.748613 \pm 0.115497$ &	N2c & $-1.342443e-05 \pm 2.753928e-06$ \\
H3c & $-1.749784 \pm 0.028088$ &	N2d & $-4.092e-4     \pm 2.847e-05$ \\
H3d & $-0.011247 \pm 0.002264$ &	N3a & $-1.659899e-04 \pm 1.788769e-05$ \\
H3e & $-0.920198 \pm 0.059910$ &	N3b & $ 5.383661e-04 \pm 2.373019e-05$ \\
H4a & $-0.434318 \pm 0.131104$ &	N3c & $ 1.238655e-03 \pm 1.918372e-05$ \\
H4b & $-1.716134 \pm 0.363024$ &	N3d & $-5.363013e-04 \pm 2.693090e-05$ \\
H4c & $ 0.619181 \pm 0.249907$ &	N4a & $ 9.409468e-04 \pm 8.455768e-05$ \\
H4d & $ 1.633127 \pm 0.195661$ &	N4b & $ 3.479228e-04 \pm 1.399697e-05$ \\
H4e & $-2.253606 \pm 0.236644$ &	N4c & $ 8.235426e-06 \pm 2.416983e-06$ \\
H4f & $-0.028194 \pm 0.041426$ &	N4d & $ 1.780791e-03 \pm 2.289154e-05$ \\
a   & $ 2.489240 \pm 0.007421$ &	N4e & $ 1.020294e-03 \pm 1.690598e-05$ \\
b   & $ 1.428658 \pm 0.035542$ &	N4f & $ 2.422e-4     \pm 6.522e-5$ \\
c   & $ 0.558505 \pm 0.052263$ &	N4g & $-7.775870e-04 \pm 6.861281e-05$ \\
    &                          &	N4h & $-5.165251e-04 \pm 1.520102e-05$ \\
    &                          &	N4i & $-1.357834e-03 \pm 6.693734e-05$ \\
\end{tabular}
\end{ruledtabular}
\end{table*}

\begin{table*}
\caption{
Table of fitting parameters for the peak frequency and amplitude of the strain 22 mode
 formulas.  Nonspinning parameters W0, A0, W2d, A2d, W4f, and A4f were determined in
Ref.~\cite{Healy:2017mvh}}\label{tab:fitparsOA}
\begin{ruledtabular}
\begin{tabular}{lr|lr}
W0  & $ 0.3587  \pm 0.0008 $ &  A0  & $ 0.3937  \pm 0.0002 $ \\
W1  & $ 0.14189 \pm 0.00009$ &	A1  & $-0.00252 \pm 0.00012$ \\
W2a & $-0.01461 \pm 0.00015$ &	A2a & $ 0.00385 \pm 0.00021$ \\
W2b & $ 0.05505 \pm 0.00023$ &	A2b & $ 0.00495 \pm 0.00031$ \\
W2c & $ 0.00878 \pm 0.00010$ &	A2c & $-0.00145 \pm 0.00012$ \\
W2d & $-0.1211  \pm 0.0036 $ &  A2d & $-0.0526  \pm 0.0015 $ \\
W3a & $-0.16841 \pm 0.00068$ &	A3a & $ 0.00331 \pm 0.00082$ \\
W3b & $ 0.04874 \pm 0.00046$ &	A3b & $ 0.01775 \pm 0.00071$ \\
W3c & $ 0.09181 \pm 0.00064$ &	A3c & $ 0.03202 \pm 0.00098$ \\
W3d & $-0.08607 \pm 0.00043$ &	A3d & $ 0.05267 \pm 0.00074$ \\
W4a & $-0.02185 \pm 0.00105$ &	A4a & $ 0.11029 \pm 0.00218$ \\
W4b & $ 0.11183 \pm 0.00047$ &	A4b & $-0.00552 \pm 0.00065$ \\
W4c & $-0.01704 \pm 0.00016$ &	A4c & $ 0.00558 \pm 0.00019$ \\
W4d & $ 0.21595 \pm 0.00138$ &	A4d & $ 0.04593 \pm 0.00211$ \\
W4e & $-0.12378 \pm 0.00090$ &	A4e & $-0.04754 \pm 0.00126$ \\
W4f & $ 0.0432  \pm 0.0034 $ &  A4f & $ 0.0179  \pm 0.0015 $ \\
W4g & $ 0.00167 \pm 0.00028$ &	A4g & $-0.00516 \pm 0.00091$ \\
W4h & $-0.13224 \pm 0.00058$ &	A4h & $ 0.00163 \pm 0.00047$ \\
W4i & $-0.09933 \pm 0.00099$ &	A4i & $-0.02098 \pm 0.00151$ \\
\end{tabular}
\end{ruledtabular}
\end{table*}

%%%%%%%%%%%%%%%%%%%%%%%%%%%%%%%%%%%%%%%%%%%%
\clearpage
\bibliographystyle{apsrev4-1}
\bibliography{../../Bibtex/references}

\end{document}